\documentclass[%
reprint,
superscriptaddress,
 amsmath,amssymb,
 aps,
]{revtex4-1}

\usepackage{graphicx}
\usepackage{dcolumn}
\usepackage{bm}

\usepackage{xcolor}
\usepackage[normalem]{ulem}

\newcommand{\blue}[1]{\textcolor{black}{#1}}

\usepackage[pagebackref=false]{hyperref}

\begin{document}

\author{L.~Banszerus}
\email{luca.banszerus@rwth-aachen.de.}
\affiliation{JARA-FIT and 2nd Institute of Physics, RWTH Aachen University, 52074 Aachen, Germany,~EU}%
\affiliation{Peter Gr\"unberg Institute  (PGI-9), Forschungszentrum J\"ulich, 52425 J\"ulich,~Germany,~EU}

\author{K.~Hecker}
\affiliation{JARA-FIT and 2nd Institute of Physics, RWTH Aachen University, 52074 Aachen, Germany,~EU}%

\author{E.~Icking}
\affiliation{JARA-FIT and 2nd Institute of Physics, RWTH Aachen University, 52074 Aachen, Germany,~EU}%
\affiliation{Peter Gr\"unberg Institute  (PGI-9), Forschungszentrum J\"ulich, 52425 J\"ulich,~Germany,~EU}

\author{S.~Trellenkamp}
\author{F.~Lentz}
\affiliation{Helmholtz Nano Facility, Forschungszentrum J\"ulich, 52425 J\"ulich,~Germany,~EU}

\author{D.~Neumaier}
\affiliation{AMO GmbH, Gesellschaft f\"ur Angewandte Mikro- und Optoelektronik, 52074 Aachen, Germany, EU}

\author{K.~Watanabe}
\affiliation{Research Center for Functional Materials, 
National Institute for Materials Science, 1-1 Namiki, Tsukuba 305-0044, Japan
}
\author{T.~Taniguchi}
\affiliation{ 
International Center for Materials Nanoarchitectonics, 
National Institute for Materials Science,  1-1 Namiki, Tsukuba 305-0044, Japan
}%

\author{C.~Volk}
\author{C.~Stampfer}
\affiliation{JARA-FIT and 2nd Institute of Physics, RWTH Aachen University, 52074 Aachen, Germany,~EU}%
\affiliation{Peter Gr\"unberg Institute  (PGI-9), Forschungszentrum J\"ulich, 52425 J\"ulich,~Germany,~EU}%

\title{Pulsed-gate spectroscopy of single-electron spin states \\ in bilayer graphene quantum dots}
\date{\today}


\begin{abstract}
Graphene and bilayer graphene quantum dots are promising hosts for spin qubits with long coherence times. 
Although recent technological improvements make it possible to confine single electrons electrostatically in bilayer graphene quantum dots, and their spin and valley texture of the single particle spectrum has been studied in detail, their relaxation dynamics remains still  
unexplored.
Here, we report on transport through a high-frequency gate controlled single-electron bilayer graphene quantum dot. By transient current spectroscopy of single-electron spin states, we extract a lower bound of the spin relaxation time 
of 0.5~$\mu$s. This result represents an important step towards the investigation of spin coherence times in graphene-based quantum dots and the implementation of spin-qubits.
\end{abstract}

\maketitle
Graphene and bilayer graphene (BLG) are  interesting materials for spintronics
thanks to the small spin-orbit coupling and the low nuclear spin density, which 
promise long spin relaxation and coherence times~\cite{Min2006Oct,Huertas-Hernando2006Oct}.
While spin relaxation in graphene has been extensively studied in recent years in the context of spin transport~\cite{Tombros2007Aug,Drogeler2016Jun,Avsar2020Jun}, very little is known about the relaxation of single-electron spins confined to a graphene or BLG quantum dot. 
This despite the fact that graphene quantum dots have received much attention, both in theory~\cite{Trauzettel2007Feb,Hachiya2014Mar,Droth2016Jan} and experiment~\cite{Ihn2010Mar,Guttinger2010Sep,Volk2011Aug,Engels2013Aug,Bischoff2012Nov,Eich2018Aug,Eich2018Jul,Banszerus2018Aug,Banszerus2020Mar,Banszerus2020Oct,Kurzmann2019Jul,Kurzmann2019Aug,Tong2020Sep}, 
as they are considered to be a promising platform for spin-based solid state quantum computation.

However, confining single electrons in graphene and BLG quantum dots, turned out to be quite challenging and, in early experiments edge disorder~\cite{Engels2013Aug,Bischoff2012Nov} prevented a precise control of the number of charge carriers and the formation of well tunable tunneling barriers. 
It is only recently, that the development of clean van der Waals heterostructures where BLG is encapsulated into hexagonal boron nitride (hBN)~\cite{Wang2013Nov} and a graphite crystal is used as a back gate~\cite{Overweg2018Jan} has allowed the realization of gate-defined quantum dots in BLG. This progress has stimulated numerous experiments demonstrating 
single-electron occupation~\cite{Eich2018Jul,Banszerus2020Mar,Banszerus2020Oct2},
excited state spectroscopy~\cite{Kurzmann2019Jul}, charge detection~\cite{Kurzmann2019Aug}, the electron-hole crossover~\cite{Banszerus2020Oct} and, most recently, tunable valley g-factors~\cite{Tong2020Sep}. However, so far no high-frequency gate control and transient current spectroscopy has been implemented in such systems and thus no dynamic processes in few-electron or single-electron BLG QDs have been investigated. High-frequency gate manipulation is a key requirement for possible spin-qubit operation, where it is used to prepare, manipulate and read-out electron spin states in quantum dots~\cite{Loss1998Jan,Watson2018Feb,Yoneda2017Dec,Nowack2011Sep}. Furthermore, it is also a versatile tool to study the dynamics and the life-times of excited states in QDs, including spin relaxation times~\cite{Fujisawa2001Feb,Fujisawa2002Sep}.

In this work, we present high-frequency, transient-current spectroscopy measurements on a single-electron quantum dot in BLG. From these measurements, we obtain a lower bound for the relaxation time of single electron spin states of 0.5~$\mu$s. 
When applying a MHz square pulse to the gate defining the QD, we observe transient currents through single-electron excited states, which have the opposite spin than the ground state. Measuring the current as a function of the pulse width, we extract characteristic blocking times, after which transient currents are suppressed. We find that the blocking times of the first three excited spin states match well with blocking processes of charge carriers tunneling from the leads into unoccupied states below the bias window, rather than relaxation processes within the QD itself, making the extracted value of $0.5$~$\mu$s merely a lower bound to the spin lifetimes of the excited states in single-electron BLG QDs. 
Our result shows both (i) that it is possible to perform pulse-gate experiments on gate-defined BLG QDs on graphitic gates, 
and (ii) that the spin relaxation time of a single electron in such edge-free nanostructures is indeed in the right order of magnitude, i.e. there are no unexpected relaxation processes with a time-scale below $0.5~\mu$s.

\begin{figure}[]
\centering
\includegraphics[draft=false,keepaspectratio=true,clip,width=\linewidth]{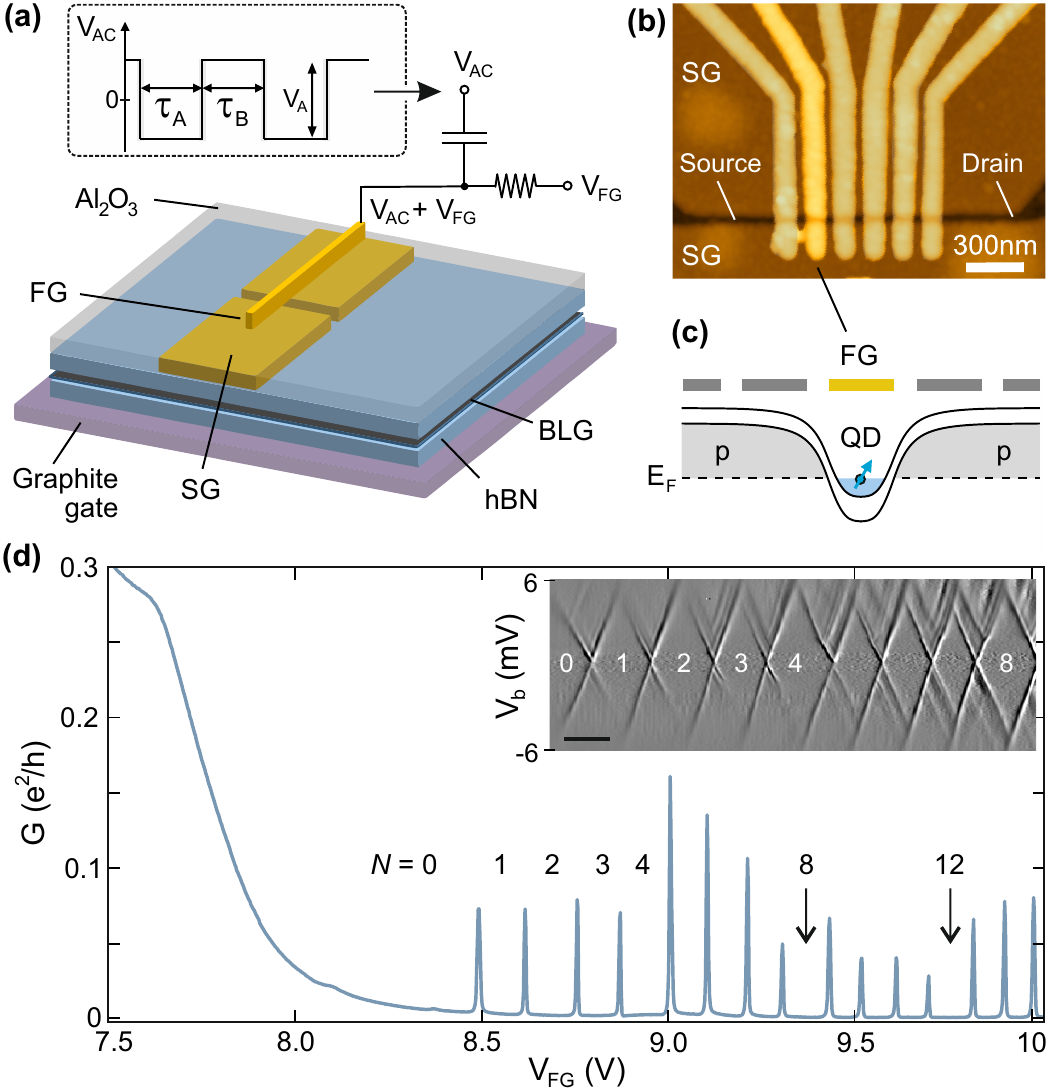}
\caption[Fig01]{
\textbf{(a)} Schematic of the device. A combination of a graphite back gate and split gates (SGs) are used to form a narrow conductive channel in the bilayer graphene. The finger gate (FG) is used to define the quantum dot and is connected to a bias-tee, allowing to apply AC pulses and DC voltages to the same gate. Inset: Illustration of a square pulse with pulse widths $\tau_\mathrm{A}$ and $\tau_\mathrm{B}$ and amplitude $V_\mathrm{A}$.
\textbf{(b)} False-color atomic force micrograph of the device showing the split gates and six finger gates (the FG used for the experiment is highlighted).
\textbf{(c)} Band edge diagram along the channel, illustrating the formation of a QD induced by a positive $V_\mathrm{FG}$ pushing the conduction band edge below the Fermi level $E_\mathrm{F}$.
\textbf{(d)} Conductance $G$ as a function of $V_\mathrm{FG}$ at a bias voltage of $V_\mathrm{b}=0.3$~mV. 
Numbers indicate the electron occupation $N$ of the QD in the regions of Coulomb blockade.
Inset: Finite bias spectroscopy $dI/dV_b$ data. The scale bar corresponds to $\Delta V_\mathrm{FG}=$100~mV.
}
\label{f1}
\end{figure}

\begin{figure}[]
\centering
\includegraphics[draft=false,keepaspectratio=true,clip,width=\linewidth]{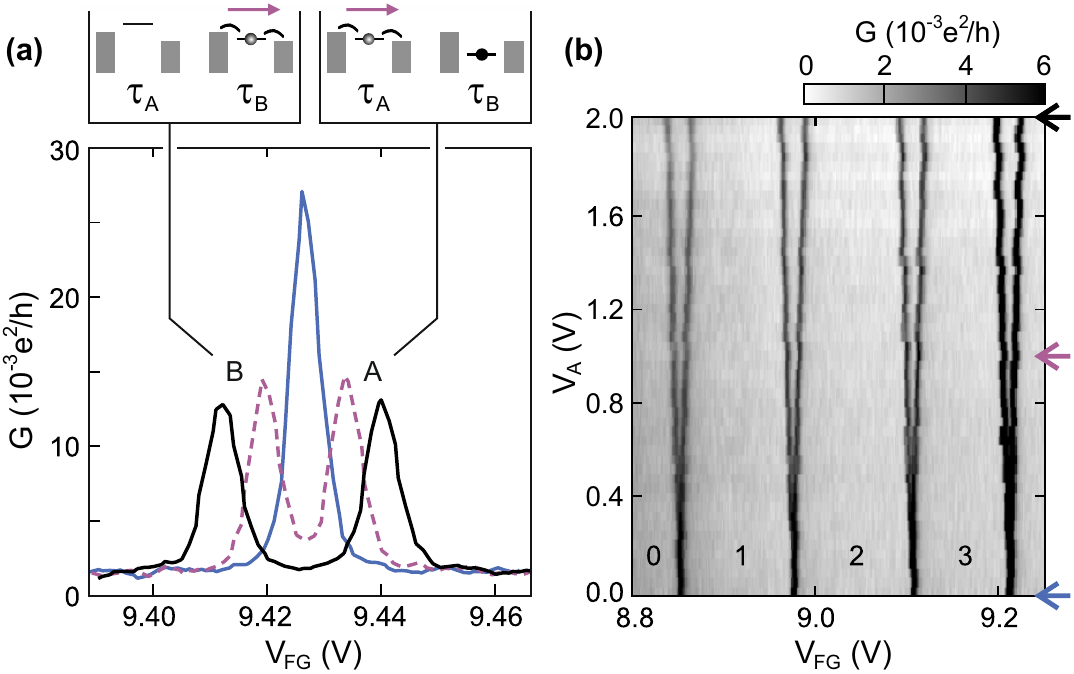}
\caption[Fig01]{
\textbf{(a)} Conductance of the 6$^\mathrm{th}$ Coulomb peak, as a function of $V_\mathrm{FG}$ while applying a square pulse of frequency $100~$kHz and $50\%$ duty cycle. Different traces correspond to AC amplitudes of $V_\mathrm{A} = $ 0~V (blue), 1~V (purple) and 2~V (black). The peak (B) at lower $V_\mathrm{FG}$ originates from tunneling events taking place during $\tau_\mathrm{B}$ and the one (A) at higher $V_\mathrm{FG}$ from tunneling events during $\tau_\mathrm{A}$ (see inset).
\textbf{(b)} Conductance as a function of $V_\mathrm{FG}$ and $V_\mathrm{A}$ for the first 4 Coulomb peaks. The arrows correspond to the $V_\mathrm{A}$ values shown in panel (a).
}
\label{f1a}
\end{figure}

A schematic illustration of the device investigated in this study is shown in Fig.~\ref{f1}(a).
The device has been fabricated from a BLG flake encapsulated between two hBN crystals of approximately 25~nm thickness using conventional van-der-Waals stacking techniques~\cite{Engels2014Sep,Wang2013Nov}. The heterostructure is placed on a graphite flake, which acts as a back gate (BG)~\cite{Banszerus2018Aug}. Cr/Au split gates with a lateral separation of 80~nm are deposited on top of the heterostructure. Isolated from the split gates by 30~nm thick atomic layer deposited Al$_2$O$_3$, we fabricate 100~nm wide finger gates (FGs) with a pitch of 150~nm to define individual quantum dots. 
Fig.~\ref{f1}(b) shows an atomic force micrograph of the gate pattern. For details of the fabrication process, we refer to Ref.~\cite{Banszerus2018Aug}. 

In order to perform pulsed gating experiments, the sample is mounted on a printed circuit board 
which is equipped with low pass filtered DC lines (1~nF capacitors to ground) as well as unfiltered 50~$\Omega$ impedance matched AC lines. 
One of the FGs is connected to a commercial bias-tee allowing for AC and DC control on the very same gate (Fig.~\ref{f1}(a)). The ohmic contacts and all other gates are connected to filtered DC lines to reduce charge noise.
All measurements are performed in a $^3$He/$^4$He dilution refrigerator at a base temperature of around 10~mK, and at an electron temperature of around 60~mK using a combination of DC-measurements and standard low-frequency lock-in techniques.

In order to form a single-electron QD, we first open a band gap in the BLG area below the split gates by applying a perpendicular electric displacement field. At a back gate voltage of $V_\mathrm{BG}=-2$~V and a split gate (SG) voltage of $V_\mathrm{SG}=1.2$~V, we induce a band gap of $\approx~25~$meV~\cite{McCann2006Mar,Oos2007Dec,Zhang2009Jun}
and fix the Fermi energy inside the gap in all regions below the SGs. 
This leaves a p-doped channel 
between the split gates, which connects source and drain (Fig.~\ref{f1}(b)). 
Subsequently, we apply a positive voltage $V_\mathrm{FG}$ to the finger gate in order to locally overcompensate the potential set by the back gate. 
This creates a p-n-p band profile along the channel (see Fig.~\ref{f1}(c)) where the Fermi energy crosses twice the band gap.
These gapped regions act as tunneling barriers separating the the QD from the reservoirs~\cite{Banszerus2018Aug,Eich2018Aug,Banszerus2020Mar}. 

\begin{figure*}[]
\centering
\includegraphics[draft=false,keepaspectratio=true,clip,width=\linewidth]{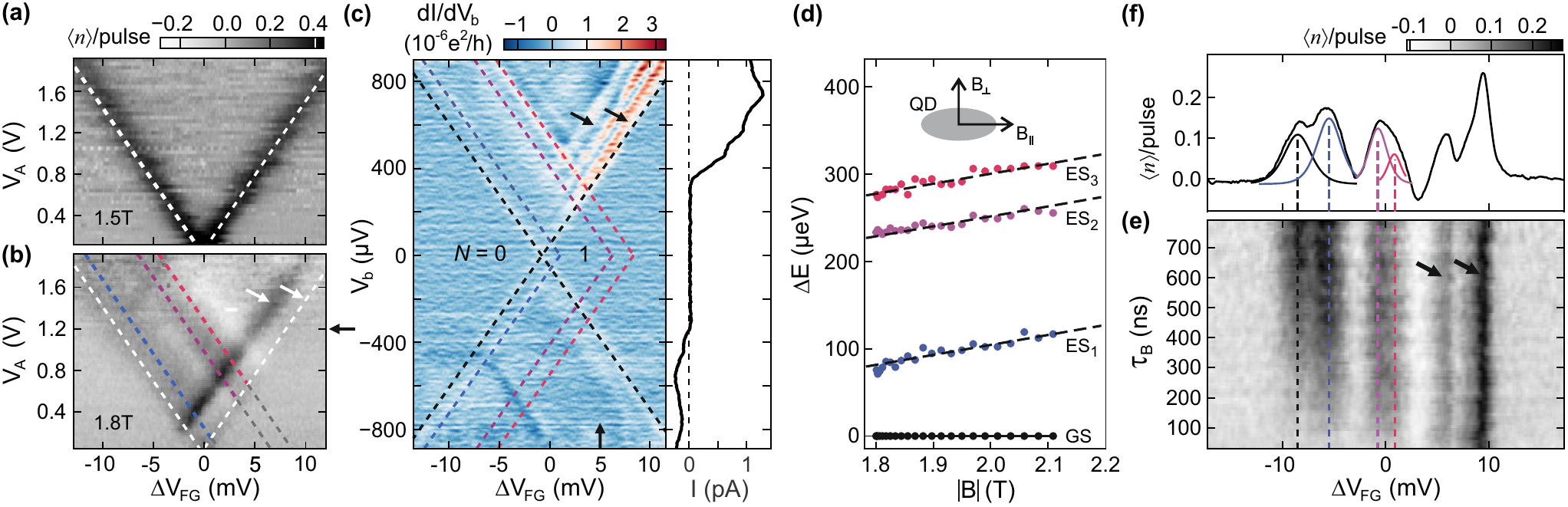}
\caption[Fig02]{\textbf{(a)} Average number of electrons $\langle n \rangle$ tunneling through the QD per pulse cycle as a function of $\Delta V_\mathrm{FG}$ and $V_\mathrm{A}$ for the 1$^{\mathrm{st}}$ Coulomb peak (transition from $N=0$ to $1$ electrons). $V_\mathrm{b}=100$~$\mathrm{\mu}$V, $B_\mathrm{\bot}=1.5$~T and $f=6.26$~MHz. Only GS transport is visible during $\tau_\mathrm{A}$ and $\tau_\mathrm{B}$ (white dashed lines). 
\textbf{(b)} Measurement as in panel (a), performed at $B_\mathrm{\bot}=1.8$~T.
Transient currents via ESs become visible. 
\textbf{(c)} Finite bias spectroscopy \blue{showing the differential conductance} in the absence of a pulsed gated voltage ($V_\mathrm{A}$=0) at the transition from $N=0$ to $1$.
The blue, purple and red dashed lines highlight the first three ESs ($B_\mathrm{\bot}=1.8$~T).
\blue{The inset shows the current along a cut at $\Delta V_\mathrm{FG} = 5$~mV, highlighting the asymmetry of the tunneling barriers~\cite{Fujisawa2001Feb}.}
\textbf{(d)} Energy difference between the different ESs and the GS measured as a function of the total magnetic field $|B|$. The in-plane magnetic field $B_\mathrm{\parallel}$ is varied at a constant $B_\mathrm{\bot}=1.8$~T. The black dashed lines correspond to the Zeeman energy shift with a g-factor of 2. 
\textbf{(e)} $\langle n \rangle$ per pulse cycle as a function of $\Delta V_\mathrm{FG}$ and $\tau_\mathrm{B}$ at $\tau_\mathrm{A}=1$~$\mathrm{\mu}s$, $V_\mathrm{A}=1.2$~V, $B_\mathrm{\bot}=1.8$~T and $B_\mathrm{\parallel}=350$~mT. \textbf{(f)} $\langle n \rangle$ per pulse cycle averaged over all $\tau_\mathrm{B}$ as a function of $\Delta V_\mathrm{FG}$. The dashed lines correspond to GS and ES marked in panels (b) and (c). }
\label{f2}
\end{figure*}

The electron occupation of the QD can be well controlled between zero and fifteen electrons by tuning $V_\mathrm{FG}$. The conductance through the device as a function of $V_\mathrm{FG}$ shown in Fig.~\ref{f1}(d).
The Coulomb peaks are arranged in groups of four, representing the fourfold shell filling pattern due to the spin and valley degeneracy in BLG in the absence of a magnetic field~\cite{Eich2018Jul,Garreis2020Nov}.   
The increase in conductance at $V_\mathrm{FG} < 8$~V arises from the formation of a continuous p-doped channel connecting the reservoirs without any tunneling barriers, as the Fermi level underneath the FG is pushed into the valence band.
In order to estimate the energy scales and the size of the formed QD, finite bias spectroscopy measurements are performed (see inset in Fig.~\ref{f1}(d)). We determine a charging energy of $E_\mathrm{C} \approx 5$~meV, corresponding to a QD capacitance of $C_\mathrm{tot}=e^2/E_\mathrm{C} \approx 32$~aF. The size of the QD is estimated by a simple plate capacitor model taking into account a BLG disk and the graphite back gate, resulting in a dot diameter of approximately 110~nm, which is in reasonable agreement with the lithographic dimensions of the device (80~nm~$\times$ 100~nm).

Next, we focus on the high frequency response of the transport through the single-electron QD.
We use an arbitrary waveform generator (Tektronix~AWG7082C) to generate an AC square-pulse with variable amplitude, $V_\mathrm{A}$, and frequency $f$. The AC signal is combined with the DC gate voltage, $V_\mathrm{FG}$, through a bias-tee at the mixing chamber of the cryostate and is applied to the FG (Fig.~\ref{f1}(a)). 
First, we apply a low frequency ($f=100$~kHz) square pulse with a duty cycle of $D=\tau_\mathrm{A}/(\tau_\mathrm{A}+\tau_\mathrm{B}) = 50\%$. Here, the variation of $V_\mathrm{AC}$ is much slower than the tunneling rates. Thus, the system can follow the pulse adiabatically resulting only in a splitting of the Coulomb peaks into two peaks of equal amplitude (see peaks labelled A and B in Fig.~\ref{f1a}(a)). 
The separation of the peaks increases linearly with the pulse amplitude $V_\mathrm{A}$ (see Figs.~\ref{f1a}(a,b)). At $V_\mathrm{A}=2$~V, the peaks are shifted by $\Delta V_\mathrm{FG}=\pm13.7$~mV relative to the Coulomb peak position at  $V_\mathrm{A}=0$~V. This peak splitting is in agreement with the attenuation of 36~dB installed on the high frequency lines of our setup. As expected, the two split peaks show approximately half the conductance compared to the Coulomb peak at $V_\mathrm{A} = 0$~V.

\begin{figure*}[]
\centering
\includegraphics[draft=false,keepaspectratio=true,clip,width=\linewidth]{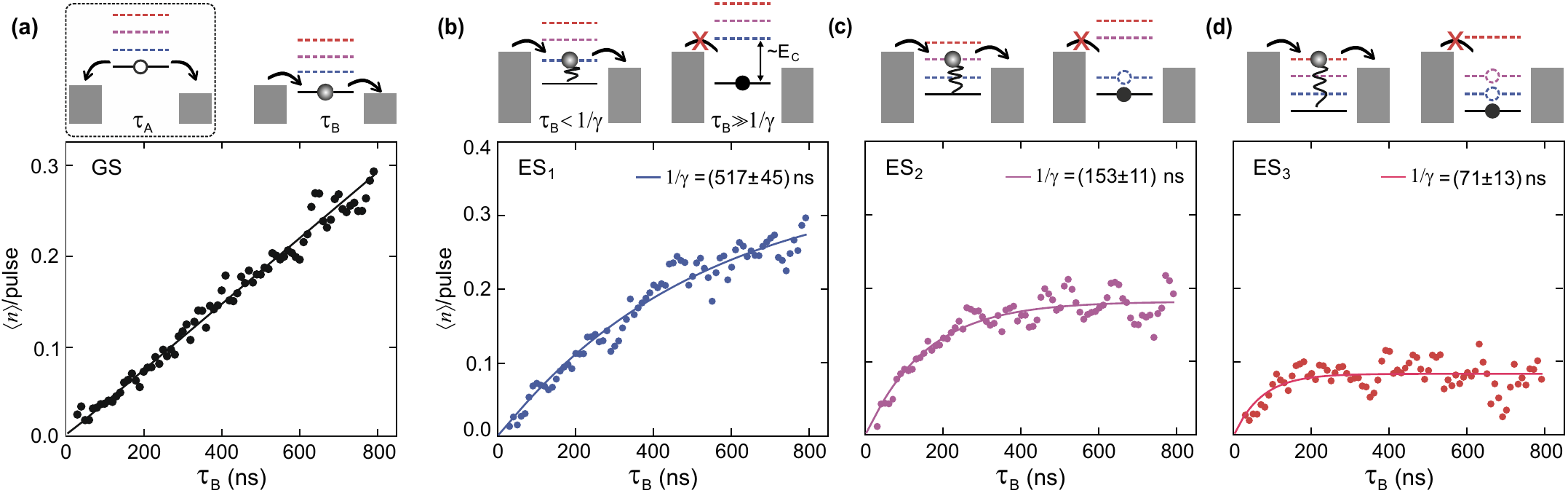}
\caption[Fig03]{\textbf{(a)} Average number of electrons $\langle n \rangle$ passing through the QD via the GS as function of $\tau_\mathrm{B}$. From the linear slope, we extract the combined tunneling rate, $\Gamma$. 
Top left: Schematic representation of the initialisation (emptying) of the QD during $\tau_A$. Top right: Possible tunneling processes during $\tau_B$ leading to a steady current through the GS. 
\textbf{(b)-(d)} Average number of electrons passing through the QD via ES$_1$, ES$_2$ and ES$_3$ as function of $\tau_\mathrm{B}$. $\langle n \rangle$/pulse saturates after the characteristic blocking time, $1/\gamma_i$. Solid lines show fits to the experimental data according to $\langle n \rangle$/pulse~$=\Gamma (1-e^{-\tau_\mathrm{B} \gamma_i})$.
Top left: Schematic representations of the possible tunneling and relaxation processes during $\tau_B$ for transport through the ES(s). Top right: Blocking of the tunneling current due to the occupation of the GS and/or ES(s) below the bias window. 
}
\label{f3}
\end{figure*}

In order to study the spin dynamics of a single-electron in the BLG QD, we now focus on the first charge transition ($N=0$ to $1$) and apply a perpendicular magnetic field, $B_\mathrm{\perp}$, to lift the spin and valley degeneracy of the QD and, crucially, to reduce the tunneling rates between the QD and the source-drain reservoirs~\cite{Eich2018Aug,Eich2018Jul,Banszerus2020Sep}. At $B_\mathrm{\perp}=1.5$~T, a tunneling current of $I \approx 720~$fA at $V_\mathrm{A}=0$~V is observed, corresponding to a combined tunneling rate through the QD of $\Gamma =I/e= \Gamma_\mathrm{L}\Gamma_\mathrm{R}/(\Gamma_\mathrm{L}+\Gamma_\mathrm{R}) \approx 4.5$~MHz. 
Note that the response time of the QD is given by the fastest tunneling rate to the leads, $\mathrm{max}(\Gamma_\mathrm{L},\Gamma_\mathrm{R})$. Fig.~\ref{f2}(a) shows the average number of electrons passing through the dot per pulse cycle, $\langle n \rangle/\mathrm{pulse} = I/e(\tau_\mathrm{A}+\tau_\mathrm{B})$, as a function of $V_\mathrm{A}$ and $\Delta V_\mathrm{FG} = V_\mathrm{FG} - V_\mathrm{FG} (V_\mathrm{A}=0)$, when applying a square pulse with a frequency of $f=6.26$~MHz. Here, the QD can still follow the applied potential adiabatically and only ground state (GS) transport is observed (white dashed lines).
By further increasing the magnetic field to $B_\mathrm{\perp}=1.8$~T, we reduce the DC tunneling current to $I\approx 60$~fA, corresponding to a combined tunneling rate of $\Gamma=0.37$~MHz. Transient currents via several excited states (ESs), e.g. ES$_1$, ES$_2$ and ES$_3$, become visible (see colored dashed lines in Fig.~\ref{f2}(b)) in addition to the ground state transport (white dashed lines). 
Features with the opposite slope correspond to ES transitions which contribute to transport during $\tau_\mathrm{A}$ (white arrows).
We compare the data with DC finite bias spectroscopy data (see Fig.~\ref{f2}(c)) in order to verify that the observed transitions match electronic excited states. Beside the charging lines of the ground state (GS), highlighted by black and white dashed lines, multiple electronic excited states can be observed as peaks in the differential conductance. 
The blue, purple and red dashed lines highlight the first three excited state transitions, observed in the transient current spectroscopy. We attribute the occurrence of this rich excited state spectrum to states of higher orbitals, which move down in energy due to the coupling of the orbital-dependent valley magnetic moment to $B_\perp$~\cite{Eich2018Aug}. 
From the bias asymmetry of the tunneling currents observed in \blue{finite bias spectroscopy (see line cut at constant gate voltage in Fig.~\ref{f2}(c)}, we estimate an asymmetry of the left and right tunneling barrier of $\Gamma_\mathrm{L}/\Gamma_\mathrm{R}\approx5$. 

To shed light on the spin texture of ES$_1$ to ES$_3$, we apply an additional in-plane magnetic field, $B_\parallel$, which exclusively couples to the spin degree of freedom. The energy of ES$_1$ to ES$_3$ as a function of the total magnetic field \blue{$|B|=\sqrt{B_\parallel^2+B_\perp^2}$ for a constant out-of-plane magnetic field $B_\perp =1.8~$T} is shown in Fig.~\ref{f2}(d). \blue{The energy differences of the ESs with respect to the GS have been obtained from finite bias spectroscopy measurements at different $B_\parallel$. The gate voltage $\Delta V_\mathrm{FG}$ has been converted into an energy scale via the lever arm $\alpha=33.6\times10^{-3}$.} 
The black dashed line corresponds to a spin g-factor of 2, which matches all three transitions, indicating that all ES transitions observed in the transient current spectroscopy require a spin flip, in order to relax into the ground state. 
The absence of ES with the same spin as the GS might be explained by fast relaxation processes into the GS, if no spin-flip is required. 

In order to determine the characteristic time scales of the transient processes, we study their dependence on the pulse width $\tau_\mathrm{B}$. Fig.~\ref{f2}(e) shows the average number of electrons $\langle n \rangle$ per pulse cycle as a function of $\Delta V_\mathrm{FG}$ and $\tau_\mathrm{B}$ at  $V_\mathrm{A}=1.2$~V (see black arrow in Fig.~\ref{f2}.(a)). As expected, $\langle n \rangle$/pulse increases with $\tau_\mathrm{B}$ for the ground state (black) and ES$_1$ to ES$_3$  (blue, purple, red), whereas the features related to $\tau_\mathrm{A}$ remain constant (black arrows). 
Fig.~\ref{f2}(f) shows $\langle n \rangle$/pulse, averaged over all $\tau_\mathrm{B}$ values shown in Fig.~3(e) as a function of $\Delta V_\mathrm{FG}$. The colored solid lines are fits to the individual excited states and highlight the different visibilities of these states.

Fig.~\ref{f3} shows cuts through Fig.~\ref{f2}(e) at constant $\Delta V_\mathrm{FG}$ (see dashed lines in Fig.~\ref{f2}(e)) as a function of $\tau_\mathrm{B}$. For transport through the GS (see Fig.~\ref{f3}(a)) no relaxation processes occur. This results in a steady state tunneling current through the dot, which is proportional to the fraction of the pulse duration, during which the GS lies in the bias window ($I \propto \tau_\mathrm{B}/(\tau_\mathrm{A}+\tau_\mathrm{B})$; see upper right schematics in Fig.~4(a)). As expected, $\langle n \rangle$/pulse increases linearly with $\tau_\mathrm{B}$. The slope is given by the combined tunneling rate of the left and right barrier resulting in $\Gamma \approx 0.36$~MHz, which is in excellent agreement with $\Gamma$ extracted from the DC current above.

Transient currents through excited states are suppressed by Coulomb blockade, once the GS or an energetically lower lying ES become occupied either by relaxation processes or by direct tunneling from the reservoirs. These processes become more probable with increasing $\tau_\mathrm{B}$, resulting in a saturation of the average number of electrons tunneling through the dot per AC pulse cycle, from which we extract the blocking time $1/\gamma$, after which the transient current is blocked (see schematics in Fig.~4(b)). Figs.~\ref{f3}(b)-4(d) show the transient currents passing the dot via ES$_1$ (blue), ES$_2$ (purple) and ES$_3$ (red) as a function of $\tau_\mathrm{B}$. In all three cases, we observe a saturation at long $\tau_\mathrm{B}$ due to the described blocking mechanisms. Following previous work, we fit a model of exponential saturation $\langle n \rangle/\mathrm{pulse}=\Gamma (1-e^{-\tau_\mathrm{B}\gamma})$ to the data, where $\gamma$ is the average blocking rate of the transient current~\cite{Volk2013Apr,Fujisawa2002Sep,Fujisawa2001Feb,Fujisawa2001Apr}. For the first ES, we observe a characteristic blocking time of $1/\gamma_1 \approx 500$~ns, while transport via the second and third ES is blocked after 150~ns and 70~ns, respectively. The higher blocking rates for the second and third ES can be understood in terms of additional mechanisms contributing to the blocking rate, if more unoccupied (excited) states reside below the bias window (see schematics in Fig.~\ref{f3}) which can be populated by tunneling from the reservoirs or \blue{spin conserving} relaxation processes. \blue{Indeed the extracted values are in good agreement with reported charge relaxation times~\cite{Volk2013Apr}.}

As the characteristic blocking time $1/\gamma_1$ is influenced by the relaxation rate of ES$_1$ to the GS, as well as by direct tunneling of electrons from the reservoirs into the GS, it provides just a lower bound for the relaxation time of ES$_1$. In fact, $1/\gamma_\mathrm{1}$ is on the same order as $1/(\Gamma_\mathrm{R}+\Gamma_\mathrm{L})$, suggesting that $\gamma_1$ is dominated by direct tunneling into the GS from source and drain. In order to reduce the influence of blocking processes due to tunneling of electrons from the reservoirs, one would need to further reduce the tunnel rates and control the asymmetry between the left and right tunneling rates. Since the observed transient currents on the order of 50~fA are close to the detection limit, this requires a capacitive readout of the tunneling events, using a charge detector.
When comparing, the extracted lower bounds for lifetimes of single-electron spin states in BLG to similar experiments in other material systems such as such as III-V semiconductor heterostructure~\cite{Hanson2003Nov} and carbon nanotubes~\cite{Churchill2009Apr}, we find that similar values on the order of 0.5~$\mu$s have been reported in these systems, all rather limited by the detection scheme.

To conclude, we have shown high-frequency gate manipulation of a single-electron BLG QD. Our transient current spectroscopy measurements of single-electron spin states give a lower bound for electronic relaxation times exceeding 0.5~$\mu$s, confirming that BLG is a promising  material for spintronics and potentially interesting for spin-based solid-state quantum computing. 
Our work motivates future experiments with better control over the dot-lead tunneling rates and integrated charge detectors enabling single-shot read-out for extracting more information on the spin dynamics in this system. Independently, the presented results is an important step towards the implementation of graphene based spin-qubits, which are now really within reach. \\\\

\textbf{Acknowledgements} This project has received funding from the European Union's Horizon 2020 research and innovation programme under grant agreement No. 881603 (Graphene Flagship) and from the European Research Council (ERC) under grant agreement No. 820254, the Deutsche Forschungsgemeinschaft (DFG, German Research Foundation) under Germany's Excellence Strategy - Cluster of Excellence Matter and Light for Quantum Computing (ML4Q) EXC 2004/1 - 390534769, through DFG (STA 1146/11-1), and by the Helmholtz Nano Facility~\cite{Albrecht2017May}. Growth of hexagonal boron nitride crystals was supported by the Elemental Strategy Initiative conducted by the MEXT, Japan, Grant Number JPMXP0112101001,  JSPS KAKENHI Grant Numbers JP20H00354 and the CREST(JPMJCR15F3), JST.

\bibliography{literature}

\begin{thebibliography}{41}%
\makeatletter
\providecommand \@ifxundefined [1]{%
 \@ifx{#1\undefined}
}%
\providecommand \@ifnum [1]{%
 \ifnum #1\expandafter \@firstoftwo
 \else \expandafter \@secondoftwo
 \fi
}%
\providecommand \@ifx [1]{%
 \ifx #1\expandafter \@firstoftwo
 \else \expandafter \@secondoftwo
 \fi
}%
\providecommand \natexlab [1]{#1}%
\providecommand \enquote  [1]{``#1''}%
\providecommand \bibnamefont  [1]{#1}%
\providecommand \bibfnamefont [1]{#1}%
\providecommand \citenamefont [1]{#1}%
\providecommand \href@noop [0]{\@secondoftwo}%
\providecommand \href [0]{\begingroup \@sanitize@url \@href}%
\providecommand \@href[1]{\@@startlink{#1}\@@href}%
\providecommand \@@href[1]{\endgroup#1\@@endlink}%
\providecommand \@sanitize@url [0]{\catcode `\\12\catcode `\$12\catcode
  `\&12\catcode `\#12\catcode `\^12\catcode `\_12\catcode `\%12\relax}%
\providecommand \@@startlink[1]{}%
\providecommand \@@endlink[0]{}%
\providecommand \url  [0]{\begingroup\@sanitize@url \@url }%
\providecommand \@url [1]{\endgroup\@href {#1}{\urlprefix }}%
\providecommand \urlprefix  [0]{URL }%
\providecommand \Eprint [0]{\href }%
\providecommand \doibase [0]{http://dx.doi.org/}%
\providecommand \selectlanguage [0]{\@gobble}%
\providecommand \bibinfo  [0]{\@secondoftwo}%
\providecommand \bibfield  [0]{\@secondoftwo}%
\providecommand \translation [1]{[#1]}%
\providecommand \BibitemOpen [0]{}%
\providecommand \bibitemStop [0]{}%
\providecommand \bibitemNoStop [0]{.\EOS\space}%
\providecommand \EOS [0]{\spacefactor3000\relax}%
\providecommand \BibitemShut  [1]{\csname bibitem#1\endcsname}%
\let\auto@bib@innerbib\@empty
\bibitem [{\citenamefont {Min}\ \emph {et~al.}(2006)\citenamefont {Min},
  \citenamefont {Hill}, \citenamefont {Sinitsyn}, \citenamefont {Sahu},
  \citenamefont {Kleinman},\ and\ \citenamefont {MacDonald}}]{Min2006Oct}%
  \BibitemOpen
  \bibfield  {author} {\bibinfo {author} {\bibfnamefont {H.}~\bibnamefont
  {Min}}, \bibinfo {author} {\bibfnamefont {J.~E.}\ \bibnamefont {Hill}},
  \bibinfo {author} {\bibfnamefont {N.~A.}\ \bibnamefont {Sinitsyn}}, \bibinfo
  {author} {\bibfnamefont {B.~R.}\ \bibnamefont {Sahu}}, \bibinfo {author}
  {\bibfnamefont {L.}~\bibnamefont {Kleinman}}, \ and\ \bibinfo {author}
  {\bibfnamefont {A.~H.}\ \bibnamefont {MacDonald}},\ }\href {\doibase
  10.1103/PhysRevB.74.165310} {\bibfield  {journal} {\bibinfo  {journal} {Phys.
  Rev. B}\ }\textbf {\bibinfo {volume} {74}},\ \bibinfo {pages} {165310}
  (\bibinfo {year} {2006})}\BibitemShut {NoStop}%
\bibitem [{\citenamefont {Huertas-Hernando}\ \emph {et~al.}(2006)\citenamefont
  {Huertas-Hernando}, \citenamefont {Guinea},\ and\ \citenamefont
  {Brataas}}]{Huertas-Hernando2006Oct}%
  \BibitemOpen
  \bibfield  {author} {\bibinfo {author} {\bibfnamefont {D.}~\bibnamefont
  {Huertas-Hernando}}, \bibinfo {author} {\bibfnamefont {F.}~\bibnamefont
  {Guinea}}, \ and\ \bibinfo {author} {\bibfnamefont {A.}~\bibnamefont
  {Brataas}},\ }\href {\doibase 10.1103/PhysRevB.74.155426} {\bibfield
  {journal} {\bibinfo  {journal} {Phys. Rev. B}\ }\textbf {\bibinfo {volume}
  {74}},\ \bibinfo {pages} {155426} (\bibinfo {year} {2006})}\BibitemShut
  {NoStop}%
\bibitem [{\citenamefont {Tombros}\ \emph {et~al.}(2007)\citenamefont
  {Tombros}, \citenamefont {Jozsa}, \citenamefont {Popinciuc}, \citenamefont
  {Jonkman},\ and\ \citenamefont {van Wees}}]{Tombros2007Aug}%
  \BibitemOpen
  \bibfield  {author} {\bibinfo {author} {\bibfnamefont {N.}~\bibnamefont
  {Tombros}}, \bibinfo {author} {\bibfnamefont {C.}~\bibnamefont {Jozsa}},
  \bibinfo {author} {\bibfnamefont {M.}~\bibnamefont {Popinciuc}}, \bibinfo
  {author} {\bibfnamefont {H.~T.}\ \bibnamefont {Jonkman}}, \ and\ \bibinfo
  {author} {\bibfnamefont {B.~J.}\ \bibnamefont {van Wees}},\ }\href {\doibase
  10.1038/nature06037} {\bibfield  {journal} {\bibinfo  {journal} {Nature}\
  }\textbf {\bibinfo {volume} {448}},\ \bibinfo {pages} {571} (\bibinfo {year}
  {2007})}\BibitemShut {NoStop}%
\bibitem [{\citenamefont {Dr{\ifmmode\ddot{o}\else\"{o}\fi}geler}\ \emph
  {et~al.}(2016)\citenamefont {Dr{\ifmmode\ddot{o}\else\"{o}\fi}geler},
  \citenamefont {Franzen}, \citenamefont {Volmer}, \citenamefont {Pohlmann},
  \citenamefont {Banszerus}, \citenamefont {Wolter}, \citenamefont {Watanabe},
  \citenamefont {Taniguchi}, \citenamefont {Stampfer},\ and\ \citenamefont
  {Beschoten}}]{Drogeler2016Jun}%
  \BibitemOpen
  \bibfield  {author} {\bibinfo {author} {\bibfnamefont {M.}~\bibnamefont
  {Dr{\ifmmode\ddot{o}\else\"{o}\fi}geler}}, \bibinfo {author} {\bibfnamefont
  {C.}~\bibnamefont {Franzen}}, \bibinfo {author} {\bibfnamefont
  {F.}~\bibnamefont {Volmer}}, \bibinfo {author} {\bibfnamefont
  {T.}~\bibnamefont {Pohlmann}}, \bibinfo {author} {\bibfnamefont
  {L.}~\bibnamefont {Banszerus}}, \bibinfo {author} {\bibfnamefont
  {M.}~\bibnamefont {Wolter}}, \bibinfo {author} {\bibfnamefont
  {K.}~\bibnamefont {Watanabe}}, \bibinfo {author} {\bibfnamefont
  {T.}~\bibnamefont {Taniguchi}}, \bibinfo {author} {\bibfnamefont
  {C.}~\bibnamefont {Stampfer}}, \ and\ \bibinfo {author} {\bibfnamefont
  {B.}~\bibnamefont {Beschoten}},\ }\href {\doibase
  10.1021/acs.nanolett.6b00497} {\bibfield  {journal} {\bibinfo  {journal}
  {Nano Lett.}\ }\textbf {\bibinfo {volume} {16}},\ \bibinfo {pages} {3533}
  (\bibinfo {year} {2016})}\BibitemShut {NoStop}%
\bibitem [{\citenamefont {Avsar}\ \emph {et~al.}(2020)\citenamefont {Avsar},
  \citenamefont {Ochoa}, \citenamefont {Guinea}, \citenamefont
  {{\ifmmode\ddot{O}\else\"{O}\fi}zyilmaz}, \citenamefont {van Wees},\ and\
  \citenamefont {Vera-Marun}}]{Avsar2020Jun}%
  \BibitemOpen
  \bibfield  {author} {\bibinfo {author} {\bibfnamefont {A.}~\bibnamefont
  {Avsar}}, \bibinfo {author} {\bibfnamefont {H.}~\bibnamefont {Ochoa}},
  \bibinfo {author} {\bibfnamefont {F.}~\bibnamefont {Guinea}}, \bibinfo
  {author} {\bibfnamefont {B.}~\bibnamefont
  {{\ifmmode\ddot{O}\else\"{O}\fi}zyilmaz}}, \bibinfo {author} {\bibfnamefont
  {B.~J.}\ \bibnamefont {van Wees}}, \ and\ \bibinfo {author} {\bibfnamefont
  {I.~J.}\ \bibnamefont {Vera-Marun}},\ }\href {\doibase
  10.1103/RevModPhys.92.021003} {\bibfield  {journal} {\bibinfo  {journal}
  {Rev. Mod. Phys.}\ }\textbf {\bibinfo {volume} {92}},\ \bibinfo {pages}
  {021003} (\bibinfo {year} {2020})}\BibitemShut {NoStop}%
\bibitem [{\citenamefont {Trauzettel}\ \emph {et~al.}(2007)\citenamefont
  {Trauzettel}, \citenamefont {Bulaev}, \citenamefont {Loss},\ and\
  \citenamefont {Burkard}}]{Trauzettel2007Feb}%
  \BibitemOpen
  \bibfield  {author} {\bibinfo {author} {\bibfnamefont {B.}~\bibnamefont
  {Trauzettel}}, \bibinfo {author} {\bibfnamefont {D.~V.}\ \bibnamefont
  {Bulaev}}, \bibinfo {author} {\bibfnamefont {D.}~\bibnamefont {Loss}}, \ and\
  \bibinfo {author} {\bibfnamefont {G.}~\bibnamefont {Burkard}},\ }\href
  {\doibase 10.1038/nphys544} {\bibfield  {journal} {\bibinfo  {journal} {Nat.
  Phys.}\ }\textbf {\bibinfo {volume} {3}},\ \bibinfo {pages} {192} (\bibinfo
  {year} {2007})}\BibitemShut {NoStop}%
\bibitem [{\citenamefont {Hachiya}\ \emph {et~al.}(2014)\citenamefont
  {Hachiya}, \citenamefont {Burkard},\ and\ \citenamefont
  {Egues}}]{Hachiya2014Mar}%
  \BibitemOpen
  \bibfield  {author} {\bibinfo {author} {\bibfnamefont {M.~O.}\ \bibnamefont
  {Hachiya}}, \bibinfo {author} {\bibfnamefont {G.}~\bibnamefont {Burkard}}, \
  and\ \bibinfo {author} {\bibfnamefont {J.~C.}\ \bibnamefont {Egues}},\ }\href
  {\doibase 10.1103/PhysRevB.89.115427} {\bibfield  {journal} {\bibinfo
  {journal} {Phys. Rev. B}\ }\textbf {\bibinfo {volume} {89}},\ \bibinfo
  {pages} {115427} (\bibinfo {year} {2014})}\BibitemShut {NoStop}%
\bibitem [{\citenamefont {Droth}\ and\ \citenamefont
  {Burkard}(2016)}]{Droth2016Jan}%
  \BibitemOpen
  \bibfield  {author} {\bibinfo {author} {\bibfnamefont {M.}~\bibnamefont
  {Droth}}\ and\ \bibinfo {author} {\bibfnamefont {G.}~\bibnamefont
  {Burkard}},\ }\href {\doibase 10.1002/pssr.201510182} {\bibfield  {journal}
  {\bibinfo  {journal} {Phys. Status Solidi RRL}\ }\textbf {\bibinfo {volume}
  {10}},\ \bibinfo {pages} {75} (\bibinfo {year} {2016})}\BibitemShut {NoStop}%
\bibitem [{\citenamefont {Ihn}\ \emph {et~al.}(2010)\citenamefont {Ihn},
  \citenamefont {G{\ifmmode \ddot{u} \else \"{u}\fi}ttinger}, \citenamefont
  {Molitor}, \citenamefont {Schnez}, \citenamefont {Schurtenberger},
  \citenamefont {Jacobsen}, \citenamefont {Hellm{\ifmmode \ddot{u} \else
  \"{u}\fi}ller}, \citenamefont {Frey}, \citenamefont {Dr{\ifmmode \ddot{o}
  \else \"{o}\fi}scher}, \citenamefont {Stampfer},\ and\ \citenamefont
  {Ensslin}}]{Ihn2010Mar}%
  \BibitemOpen
  \bibfield  {author} {\bibinfo {author} {\bibfnamefont {T.}~\bibnamefont
  {Ihn}}, \bibinfo {author} {\bibfnamefont {J.}~\bibnamefont {G{\ifmmode
  \ddot{u} \else \"{u}\fi}ttinger}}, \bibinfo {author} {\bibfnamefont
  {F.}~\bibnamefont {Molitor}}, \bibinfo {author} {\bibfnamefont
  {S.}~\bibnamefont {Schnez}}, \bibinfo {author} {\bibfnamefont
  {E.}~\bibnamefont {Schurtenberger}}, \bibinfo {author} {\bibfnamefont
  {A.}~\bibnamefont {Jacobsen}}, \bibinfo {author} {\bibfnamefont
  {S.}~\bibnamefont {Hellm{\ifmmode \ddot{u} \else \"{u}\fi}ller}}, \bibinfo
  {author} {\bibfnamefont {T.}~\bibnamefont {Frey}}, \bibinfo {author}
  {\bibfnamefont {S.}~\bibnamefont {Dr{\ifmmode \ddot{o} \else
  \"{o}\fi}scher}}, \bibinfo {author} {\bibfnamefont {C.}~\bibnamefont
  {Stampfer}}, \ and\ \bibinfo {author} {\bibfnamefont {K.}~\bibnamefont
  {Ensslin}},\ }\href {\doibase 10.1016/S1369-7021(10)70033-X} {\bibfield
  {journal} {\bibinfo  {journal} {Mater. Today}\ }\textbf {\bibinfo {volume}
  {13}},\ \bibinfo {pages} {44} (\bibinfo {year} {2010})}\BibitemShut {NoStop}%
\bibitem [{\citenamefont {G{\ifmmode \ddot{u} \else \"{u}\fi}ttinger}\ \emph
  {et~al.}(2010)\citenamefont {G{\ifmmode \ddot{u} \else \"{u}\fi}ttinger},
  \citenamefont {Frey}, \citenamefont {Stampfer}, \citenamefont {Ihn},\ and\
  \citenamefont {Ensslin}}]{Guttinger2010Sep}%
  \BibitemOpen
  \bibfield  {author} {\bibinfo {author} {\bibfnamefont {J.}~\bibnamefont
  {G{\ifmmode \ddot{u} \else \"{u}\fi}ttinger}}, \bibinfo {author}
  {\bibfnamefont {T.}~\bibnamefont {Frey}}, \bibinfo {author} {\bibfnamefont
  {C.}~\bibnamefont {Stampfer}}, \bibinfo {author} {\bibfnamefont
  {T.}~\bibnamefont {Ihn}}, \ and\ \bibinfo {author} {\bibfnamefont
  {K.}~\bibnamefont {Ensslin}},\ }\href {\doibase
  10.1103/PhysRevLett.105.116801} {\bibfield  {journal} {\bibinfo  {journal}
  {Phys. Rev. Lett.}\ }\textbf {\bibinfo {volume} {105}},\ \bibinfo {pages}
  {116801} (\bibinfo {year} {2010})}\BibitemShut {NoStop}%
\bibitem [{\citenamefont {Volk}\ \emph {et~al.}(2011)\citenamefont {Volk},
  \citenamefont {Fringes}, \citenamefont {Terr{\ifmmode \acute{e} \else
  \'{e}\fi}s}, \citenamefont {Dauber}, \citenamefont {Engels}, \citenamefont
  {Trellenkamp},\ and\ \citenamefont {Stampfer}}]{Volk2011Aug}%
  \BibitemOpen
  \bibfield  {author} {\bibinfo {author} {\bibfnamefont {C.}~\bibnamefont
  {Volk}}, \bibinfo {author} {\bibfnamefont {S.}~\bibnamefont {Fringes}},
  \bibinfo {author} {\bibfnamefont {B.}~\bibnamefont {Terr{\ifmmode \acute{e}
  \else \'{e}\fi}s}}, \bibinfo {author} {\bibfnamefont {J.}~\bibnamefont
  {Dauber}}, \bibinfo {author} {\bibfnamefont {S.}~\bibnamefont {Engels}},
  \bibinfo {author} {\bibfnamefont {S.}~\bibnamefont {Trellenkamp}}, \ and\
  \bibinfo {author} {\bibfnamefont {C.}~\bibnamefont {Stampfer}},\ }\href
  {\doibase 10.1021/nl201295s} {\bibfield  {journal} {\bibinfo  {journal} {Nano
  Lett.}\ }\textbf {\bibinfo {volume} {11}},\ \bibinfo {pages} {3581} (\bibinfo
  {year} {2011})}\BibitemShut {NoStop}%
\bibitem [{\citenamefont {Engels}\ \emph {et~al.}(2013)\citenamefont {Engels},
  \citenamefont {Epping}, \citenamefont {Volk}, \citenamefont {Korte},
  \citenamefont {Voigtl{\ifmmode \ddot{a} \else \"{a}\fi}nder}, \citenamefont
  {Watanabe}, \citenamefont {Taniguchi}, \citenamefont {Trellenkamp},\ and\
  \citenamefont {Stampfer}}]{Engels2013Aug}%
  \BibitemOpen
  \bibfield  {author} {\bibinfo {author} {\bibfnamefont {S.}~\bibnamefont
  {Engels}}, \bibinfo {author} {\bibfnamefont {A.}~\bibnamefont {Epping}},
  \bibinfo {author} {\bibfnamefont {C.}~\bibnamefont {Volk}}, \bibinfo {author}
  {\bibfnamefont {S.}~\bibnamefont {Korte}}, \bibinfo {author} {\bibfnamefont
  {B.}~\bibnamefont {Voigtl{\ifmmode \ddot{a} \else \"{a}\fi}nder}}, \bibinfo
  {author} {\bibfnamefont {K.}~\bibnamefont {Watanabe}}, \bibinfo {author}
  {\bibfnamefont {T.}~\bibnamefont {Taniguchi}}, \bibinfo {author}
  {\bibfnamefont {S.}~\bibnamefont {Trellenkamp}}, \ and\ \bibinfo {author}
  {\bibfnamefont {C.}~\bibnamefont {Stampfer}},\ }\href {\doibase
  10.1063/1.4818627} {\bibfield  {journal} {\bibinfo  {journal} {Appl. Phys.
  Lett.}\ }\textbf {\bibinfo {volume} {103}},\ \bibinfo {pages} {073113}
  (\bibinfo {year} {2013})}\BibitemShut {NoStop}%
\bibitem [{\citenamefont {Bischoff}\ \emph {et~al.}(2012)\citenamefont
  {Bischoff}, \citenamefont {Kr{\ifmmode \ddot{a} \else \"{a}\fi}henmann},
  \citenamefont {Dr{\ifmmode \ddot{o} \else \"{o}\fi}scher}, \citenamefont
  {Gruner}, \citenamefont {Barraud}, \citenamefont {Ihn},\ and\ \citenamefont
  {Ensslin}}]{Bischoff2012Nov}%
  \BibitemOpen
  \bibfield  {author} {\bibinfo {author} {\bibfnamefont {D.}~\bibnamefont
  {Bischoff}}, \bibinfo {author} {\bibfnamefont {T.}~\bibnamefont {Kr{\ifmmode
  \ddot{a} \else \"{a}\fi}henmann}}, \bibinfo {author} {\bibfnamefont
  {S.}~\bibnamefont {Dr{\ifmmode \ddot{o} \else \"{o}\fi}scher}}, \bibinfo
  {author} {\bibfnamefont {M.~A.}\ \bibnamefont {Gruner}}, \bibinfo {author}
  {\bibfnamefont {C.}~\bibnamefont {Barraud}}, \bibinfo {author} {\bibfnamefont
  {T.}~\bibnamefont {Ihn}}, \ and\ \bibinfo {author} {\bibfnamefont
  {K.}~\bibnamefont {Ensslin}},\ }\href {\doibase 10.1063/1.4765345} {\bibfield
   {journal} {\bibinfo  {journal} {Appl. Phys. Lett.}\ }\textbf {\bibinfo
  {volume} {101}},\ \bibinfo {pages} {203103} (\bibinfo {year}
  {2012})}\BibitemShut {NoStop}%
\bibitem [{\citenamefont {Eich}\ \emph
  {et~al.}(2018{\natexlab{a}})\citenamefont {Eich}, \citenamefont {Pisoni},
  \citenamefont {Pally}, \citenamefont {Overweg}, \citenamefont {Kurzmann},
  \citenamefont {Lee}, \citenamefont {Rickhaus}, \citenamefont {Watanabe},
  \citenamefont {Taniguchi}, \citenamefont {Ensslin},\ and\ \citenamefont
  {Ihn}}]{Eich2018Aug}%
  \BibitemOpen
  \bibfield  {author} {\bibinfo {author} {\bibfnamefont {M.}~\bibnamefont
  {Eich}}, \bibinfo {author} {\bibfnamefont {R.}~\bibnamefont {Pisoni}},
  \bibinfo {author} {\bibfnamefont {A.}~\bibnamefont {Pally}}, \bibinfo
  {author} {\bibfnamefont {H.}~\bibnamefont {Overweg}}, \bibinfo {author}
  {\bibfnamefont {A.}~\bibnamefont {Kurzmann}}, \bibinfo {author}
  {\bibfnamefont {Y.}~\bibnamefont {Lee}}, \bibinfo {author} {\bibfnamefont
  {P.}~\bibnamefont {Rickhaus}}, \bibinfo {author} {\bibfnamefont
  {K.}~\bibnamefont {Watanabe}}, \bibinfo {author} {\bibfnamefont
  {T.}~\bibnamefont {Taniguchi}}, \bibinfo {author} {\bibfnamefont
  {K.}~\bibnamefont {Ensslin}}, \ and\ \bibinfo {author} {\bibfnamefont
  {T.}~\bibnamefont {Ihn}},\ }\href {\doibase 10.1021/acs.nanolett.8b01859}
  {\bibfield  {journal} {\bibinfo  {journal} {Nano Lett.}\ }\textbf {\bibinfo
  {volume} {18}},\ \bibinfo {pages} {5042} (\bibinfo {year}
  {2018}{\natexlab{a}})}\BibitemShut {NoStop}%
\bibitem [{\citenamefont {Eich}\ \emph
  {et~al.}(2018{\natexlab{b}})\citenamefont {Eich}, \citenamefont {Pisoni},
  \citenamefont {Overweg}, \citenamefont {Kurzmann}, \citenamefont {Lee},
  \citenamefont {Rickhaus}, \citenamefont {Ihn}, \citenamefont {Ensslin},
  \citenamefont {Herman}, \citenamefont {Sigrist}, \citenamefont {Watanabe},\
  and\ \citenamefont {Taniguchi}}]{Eich2018Jul}%
  \BibitemOpen
  \bibfield  {author} {\bibinfo {author} {\bibfnamefont {M.}~\bibnamefont
  {Eich}}, \bibinfo {author} {\bibfnamefont {R.}~\bibnamefont {Pisoni}},
  \bibinfo {author} {\bibfnamefont {H.}~\bibnamefont {Overweg}}, \bibinfo
  {author} {\bibfnamefont {A.}~\bibnamefont {Kurzmann}}, \bibinfo {author}
  {\bibfnamefont {Y.}~\bibnamefont {Lee}}, \bibinfo {author} {\bibfnamefont
  {P.}~\bibnamefont {Rickhaus}}, \bibinfo {author} {\bibfnamefont
  {T.}~\bibnamefont {Ihn}}, \bibinfo {author} {\bibfnamefont {K.}~\bibnamefont
  {Ensslin}}, \bibinfo {author} {\bibfnamefont {F.}~\bibnamefont {Herman}},
  \bibinfo {author} {\bibfnamefont {M.}~\bibnamefont {Sigrist}}, \bibinfo
  {author} {\bibfnamefont {K.}~\bibnamefont {Watanabe}}, \ and\ \bibinfo
  {author} {\bibfnamefont {T.}~\bibnamefont {Taniguchi}},\ }\href {\doibase
  10.1103/PhysRevX.8.031023} {\bibfield  {journal} {\bibinfo  {journal} {Phys.
  Rev. X}\ }\textbf {\bibinfo {volume} {8}},\ \bibinfo {pages} {031023}
  (\bibinfo {year} {2018}{\natexlab{b}})}\BibitemShut {NoStop}%
\bibitem [{\citenamefont {Banszerus}\ \emph {et~al.}(2018)\citenamefont
  {Banszerus}, \citenamefont {Frohn}, \citenamefont {Epping}, \citenamefont
  {Neumaier}, \citenamefont {Watanabe}, \citenamefont {Taniguchi},\ and\
  \citenamefont {Stampfer}}]{Banszerus2018Aug}%
  \BibitemOpen
  \bibfield  {author} {\bibinfo {author} {\bibfnamefont {L.}~\bibnamefont
  {Banszerus}}, \bibinfo {author} {\bibfnamefont {B.}~\bibnamefont {Frohn}},
  \bibinfo {author} {\bibfnamefont {A.}~\bibnamefont {Epping}}, \bibinfo
  {author} {\bibfnamefont {D.}~\bibnamefont {Neumaier}}, \bibinfo {author}
  {\bibfnamefont {K.}~\bibnamefont {Watanabe}}, \bibinfo {author}
  {\bibfnamefont {T.}~\bibnamefont {Taniguchi}}, \ and\ \bibinfo {author}
  {\bibfnamefont {C.}~\bibnamefont {Stampfer}},\ }\href {\doibase
  10.1021/acs.nanolett.8b01303} {\bibfield  {journal} {\bibinfo  {journal}
  {Nano Lett.}\ }\textbf {\bibinfo {volume} {18}},\ \bibinfo {pages} {4785}
  (\bibinfo {year} {2018})}\BibitemShut {NoStop}%
\bibitem [{\citenamefont {Banszerus}\ \emph
  {et~al.}(2020{\natexlab{a}})\citenamefont {Banszerus}, \citenamefont
  {M{\ifmmode\ddot{o}\else\"{o}\fi}ller}, \citenamefont {Icking}, \citenamefont
  {Watanabe}, \citenamefont {Taniguchi}, \citenamefont {Volk},\ and\
  \citenamefont {Stampfer}}]{Banszerus2020Mar}%
  \BibitemOpen
  \bibfield  {author} {\bibinfo {author} {\bibfnamefont {L.}~\bibnamefont
  {Banszerus}}, \bibinfo {author} {\bibfnamefont {S.}~\bibnamefont
  {M{\ifmmode\ddot{o}\else\"{o}\fi}ller}}, \bibinfo {author} {\bibfnamefont
  {E.}~\bibnamefont {Icking}}, \bibinfo {author} {\bibfnamefont
  {K.}~\bibnamefont {Watanabe}}, \bibinfo {author} {\bibfnamefont
  {T.}~\bibnamefont {Taniguchi}}, \bibinfo {author} {\bibfnamefont
  {C.}~\bibnamefont {Volk}}, \ and\ \bibinfo {author} {\bibfnamefont
  {C.}~\bibnamefont {Stampfer}},\ }\href {\doibase
  10.1021/acs.nanolett.9b05295} {\bibfield  {journal} {\bibinfo  {journal}
  {Nano Lett.}\ }\textbf {\bibinfo {volume} {20}},\ \bibinfo {pages} {2005}
  (\bibinfo {year} {2020}{\natexlab{a}})}\BibitemShut {NoStop}%
\bibitem [{\citenamefont {Banszerus}\ \emph
  {et~al.}(2020{\natexlab{b}})\citenamefont {Banszerus}, \citenamefont
  {Rothstein}, \citenamefont {Fabian}, \citenamefont
  {M{\ifmmode\ddot{o}\else\"{o}\fi}ller}, \citenamefont {Icking}, \citenamefont
  {Trellenkamp}, \citenamefont {Lentz}, \citenamefont {Neumaier}, \citenamefont
  {Watanabe}, \citenamefont {Taniguchi}, \citenamefont {Libisch}, \citenamefont
  {Volk},\ and\ \citenamefont {Stampfer}}]{Banszerus2020Oct}%
  \BibitemOpen
  \bibfield  {author} {\bibinfo {author} {\bibfnamefont {L.}~\bibnamefont
  {Banszerus}}, \bibinfo {author} {\bibfnamefont {A.}~\bibnamefont
  {Rothstein}}, \bibinfo {author} {\bibfnamefont {T.}~\bibnamefont {Fabian}},
  \bibinfo {author} {\bibfnamefont {S.}~\bibnamefont
  {M{\ifmmode\ddot{o}\else\"{o}\fi}ller}}, \bibinfo {author} {\bibfnamefont
  {E.}~\bibnamefont {Icking}}, \bibinfo {author} {\bibfnamefont
  {S.}~\bibnamefont {Trellenkamp}}, \bibinfo {author} {\bibfnamefont
  {F.}~\bibnamefont {Lentz}}, \bibinfo {author} {\bibfnamefont
  {D.}~\bibnamefont {Neumaier}}, \bibinfo {author} {\bibfnamefont
  {K.}~\bibnamefont {Watanabe}}, \bibinfo {author} {\bibfnamefont
  {T.}~\bibnamefont {Taniguchi}}, \bibinfo {author} {\bibfnamefont
  {F.}~\bibnamefont {Libisch}}, \bibinfo {author} {\bibfnamefont
  {C.}~\bibnamefont {Volk}}, \ and\ \bibinfo {author} {\bibfnamefont
  {C.}~\bibnamefont {Stampfer}},\ }\href {\doibase
  10.1021/acs.nanolett.0c03227} {\bibfield  {journal} {\bibinfo  {journal}
  {Nano Lett.}\ }\textbf {\bibinfo {volume} {20}},\ \bibinfo {pages} {7709}
  (\bibinfo {year} {2020}{\natexlab{b}})}\BibitemShut {NoStop}%
\bibitem [{\citenamefont {Kurzmann}\ \emph
  {et~al.}(2019{\natexlab{a}})\citenamefont {Kurzmann}, \citenamefont {Eich},
  \citenamefont {Overweg}, \citenamefont {Mangold}, \citenamefont {Herman},
  \citenamefont {Rickhaus}, \citenamefont {Pisoni}, \citenamefont {Lee},
  \citenamefont {Garreis}, \citenamefont {Tong}, \citenamefont {Watanabe},
  \citenamefont {Taniguchi}, \citenamefont {Ensslin},\ and\ \citenamefont
  {Ihn}}]{Kurzmann2019Jul}%
  \BibitemOpen
  \bibfield  {author} {\bibinfo {author} {\bibfnamefont {A.}~\bibnamefont
  {Kurzmann}}, \bibinfo {author} {\bibfnamefont {M.}~\bibnamefont {Eich}},
  \bibinfo {author} {\bibfnamefont {H.}~\bibnamefont {Overweg}}, \bibinfo
  {author} {\bibfnamefont {M.}~\bibnamefont {Mangold}}, \bibinfo {author}
  {\bibfnamefont {F.}~\bibnamefont {Herman}}, \bibinfo {author} {\bibfnamefont
  {P.}~\bibnamefont {Rickhaus}}, \bibinfo {author} {\bibfnamefont
  {R.}~\bibnamefont {Pisoni}}, \bibinfo {author} {\bibfnamefont
  {Y.}~\bibnamefont {Lee}}, \bibinfo {author} {\bibfnamefont {R.}~\bibnamefont
  {Garreis}}, \bibinfo {author} {\bibfnamefont {C.}~\bibnamefont {Tong}},
  \bibinfo {author} {\bibfnamefont {K.}~\bibnamefont {Watanabe}}, \bibinfo
  {author} {\bibfnamefont {T.}~\bibnamefont {Taniguchi}}, \bibinfo {author}
  {\bibfnamefont {K.}~\bibnamefont {Ensslin}}, \ and\ \bibinfo {author}
  {\bibfnamefont {T.}~\bibnamefont {Ihn}},\ }\href {\doibase
  10.1103/PhysRevLett.123.026803} {\bibfield  {journal} {\bibinfo  {journal}
  {Phys. Rev. Lett.}\ }\textbf {\bibinfo {volume} {123}},\ \bibinfo {pages}
  {026803} (\bibinfo {year} {2019}{\natexlab{a}})}\BibitemShut {NoStop}%
\bibitem [{\citenamefont {Kurzmann}\ \emph
  {et~al.}(2019{\natexlab{b}})\citenamefont {Kurzmann}, \citenamefont
  {Overweg}, \citenamefont {Eich}, \citenamefont {Pally}, \citenamefont
  {Rickhaus}, \citenamefont {Pisoni}, \citenamefont {Lee}, \citenamefont
  {Watanabe}, \citenamefont {Taniguchi}, \citenamefont {Ihn},\ and\
  \citenamefont {Ensslin}}]{Kurzmann2019Aug}%
  \BibitemOpen
  \bibfield  {author} {\bibinfo {author} {\bibfnamefont {A.}~\bibnamefont
  {Kurzmann}}, \bibinfo {author} {\bibfnamefont {H.}~\bibnamefont {Overweg}},
  \bibinfo {author} {\bibfnamefont {M.}~\bibnamefont {Eich}}, \bibinfo {author}
  {\bibfnamefont {A.}~\bibnamefont {Pally}}, \bibinfo {author} {\bibfnamefont
  {P.}~\bibnamefont {Rickhaus}}, \bibinfo {author} {\bibfnamefont
  {R.}~\bibnamefont {Pisoni}}, \bibinfo {author} {\bibfnamefont
  {Y.}~\bibnamefont {Lee}}, \bibinfo {author} {\bibfnamefont {K.}~\bibnamefont
  {Watanabe}}, \bibinfo {author} {\bibfnamefont {T.}~\bibnamefont {Taniguchi}},
  \bibinfo {author} {\bibfnamefont {T.}~\bibnamefont {Ihn}}, \ and\ \bibinfo
  {author} {\bibfnamefont {K.}~\bibnamefont {Ensslin}},\ }\href {\doibase
  10.1021/acs.nanolett.9b01617} {\bibfield  {journal} {\bibinfo  {journal}
  {Nano Lett.}\ }\textbf {\bibinfo {volume} {19}},\ \bibinfo {pages} {5216}
  (\bibinfo {year} {2019}{\natexlab{b}})}\BibitemShut {NoStop}%
\bibitem [{\citenamefont {Tong}\ \emph {et~al.}(2020)\citenamefont {Tong},
  \citenamefont {Garreis}, \citenamefont {Knothe}, \citenamefont {Eich},
  \citenamefont {Sacchi}, \citenamefont {Watanabe}, \citenamefont {Taniguchi},
  \citenamefont {Fal'ko}, \citenamefont {Ihn}, \citenamefont {Ensslin},\ and\
  \citenamefont {Kurzmann}}]{Tong2020Sep}%
  \BibitemOpen
  \bibfield  {author} {\bibinfo {author} {\bibfnamefont {C.}~\bibnamefont
  {Tong}}, \bibinfo {author} {\bibfnamefont {R.}~\bibnamefont {Garreis}},
  \bibinfo {author} {\bibfnamefont {A.}~\bibnamefont {Knothe}}, \bibinfo
  {author} {\bibfnamefont {M.}~\bibnamefont {Eich}}, \bibinfo {author}
  {\bibfnamefont {A.}~\bibnamefont {Sacchi}}, \bibinfo {author} {\bibfnamefont
  {K.}~\bibnamefont {Watanabe}}, \bibinfo {author} {\bibfnamefont
  {T.}~\bibnamefont {Taniguchi}}, \bibinfo {author} {\bibfnamefont
  {V.}~\bibnamefont {Fal'ko}}, \bibinfo {author} {\bibfnamefont
  {T.}~\bibnamefont {Ihn}}, \bibinfo {author} {\bibfnamefont {K.}~\bibnamefont
  {Ensslin}}, \ and\ \bibinfo {author} {\bibfnamefont {A.}~\bibnamefont
  {Kurzmann}},\ }\href {https://arxiv.org/abs/2009.04337v1} {\bibfield
  {journal} {\bibinfo  {journal} {arXiv}\ } (\bibinfo {year} {2020})},\ \Eprint
  {http://arxiv.org/abs/2009.04337} {2009.04337} \BibitemShut {NoStop}%
\bibitem [{\citenamefont {Wang}\ \emph {et~al.}(2013)\citenamefont {Wang},
  \citenamefont {Meric}, \citenamefont {Huang}, \citenamefont {Gao},
  \citenamefont {Gao}, \citenamefont {Tran}, \citenamefont {Taniguchi},
  \citenamefont {Watanabe}, \citenamefont {Campos}, \citenamefont {Muller},
  \citenamefont {Guo}, \citenamefont {Kim}, \citenamefont {Hone}, \citenamefont
  {Shepard},\ and\ \citenamefont {Dean}}]{Wang2013Nov}%
  \BibitemOpen
  \bibfield  {author} {\bibinfo {author} {\bibfnamefont {L.}~\bibnamefont
  {Wang}}, \bibinfo {author} {\bibfnamefont {I.}~\bibnamefont {Meric}},
  \bibinfo {author} {\bibfnamefont {P.~Y.}\ \bibnamefont {Huang}}, \bibinfo
  {author} {\bibfnamefont {Q.}~\bibnamefont {Gao}}, \bibinfo {author}
  {\bibfnamefont {Y.}~\bibnamefont {Gao}}, \bibinfo {author} {\bibfnamefont
  {H.}~\bibnamefont {Tran}}, \bibinfo {author} {\bibfnamefont {T.}~\bibnamefont
  {Taniguchi}}, \bibinfo {author} {\bibfnamefont {K.}~\bibnamefont {Watanabe}},
  \bibinfo {author} {\bibfnamefont {L.~M.}\ \bibnamefont {Campos}}, \bibinfo
  {author} {\bibfnamefont {D.~A.}\ \bibnamefont {Muller}}, \bibinfo {author}
  {\bibfnamefont {J.}~\bibnamefont {Guo}}, \bibinfo {author} {\bibfnamefont
  {P.}~\bibnamefont {Kim}}, \bibinfo {author} {\bibfnamefont {J.}~\bibnamefont
  {Hone}}, \bibinfo {author} {\bibfnamefont {K.~L.}\ \bibnamefont {Shepard}}, \
  and\ \bibinfo {author} {\bibfnamefont {C.~R.}\ \bibnamefont {Dean}},\ }\href
  {\doibase 10.1126/science.1244358} {\bibfield  {journal} {\bibinfo  {journal}
  {Science}\ }\textbf {\bibinfo {volume} {342}},\ \bibinfo {pages} {614}
  (\bibinfo {year} {2013})}\BibitemShut {NoStop}%
\bibitem [{\citenamefont {Overweg}\ \emph {et~al.}(2018)\citenamefont
  {Overweg}, \citenamefont {Eggimann}, \citenamefont {Chen}, \citenamefont
  {Slizovskiy}, \citenamefont {Eich}, \citenamefont {Pisoni}, \citenamefont
  {Lee}, \citenamefont {Rickhaus}, \citenamefont {Watanabe}, \citenamefont
  {Taniguchi}, \citenamefont {Fal{'}ko}, \citenamefont {Ihn},\ and\
  \citenamefont {Ensslin}}]{Overweg2018Jan}%
  \BibitemOpen
  \bibfield  {author} {\bibinfo {author} {\bibfnamefont {H.}~\bibnamefont
  {Overweg}}, \bibinfo {author} {\bibfnamefont {H.}~\bibnamefont {Eggimann}},
  \bibinfo {author} {\bibfnamefont {X.}~\bibnamefont {Chen}}, \bibinfo {author}
  {\bibfnamefont {S.}~\bibnamefont {Slizovskiy}}, \bibinfo {author}
  {\bibfnamefont {M.}~\bibnamefont {Eich}}, \bibinfo {author} {\bibfnamefont
  {R.}~\bibnamefont {Pisoni}}, \bibinfo {author} {\bibfnamefont
  {Y.}~\bibnamefont {Lee}}, \bibinfo {author} {\bibfnamefont {P.}~\bibnamefont
  {Rickhaus}}, \bibinfo {author} {\bibfnamefont {K.}~\bibnamefont {Watanabe}},
  \bibinfo {author} {\bibfnamefont {T.}~\bibnamefont {Taniguchi}}, \bibinfo
  {author} {\bibfnamefont {V.}~\bibnamefont {Fal{'}ko}}, \bibinfo {author}
  {\bibfnamefont {T.}~\bibnamefont {Ihn}}, \ and\ \bibinfo {author}
  {\bibfnamefont {K.}~\bibnamefont {Ensslin}},\ }\href {\doibase
  10.1021/acs.nanolett.7b04666} {\bibfield  {journal} {\bibinfo  {journal}
  {Nano Lett.}\ }\textbf {\bibinfo {volume} {18}},\ \bibinfo {pages} {553}
  (\bibinfo {year} {2018})}\BibitemShut {NoStop}%
\bibitem [{\citenamefont {Banszerus}\ \emph
  {et~al.}(2020{\natexlab{c}})\citenamefont {Banszerus}, \citenamefont
  {Rothstein}, \citenamefont {Icking}, \citenamefont
  {M{\ifmmode\ddot{o}\else\"{o}\fi}ller}, \citenamefont {Watanabe},
  \citenamefont {Taniguchi}, \citenamefont {Stampfer},\ and\ \citenamefont
  {Volk}}]{Banszerus2020Oct2}%
  \BibitemOpen
  \bibfield  {author} {\bibinfo {author} {\bibfnamefont {L.}~\bibnamefont
  {Banszerus}}, \bibinfo {author} {\bibfnamefont {A.}~\bibnamefont
  {Rothstein}}, \bibinfo {author} {\bibfnamefont {E.}~\bibnamefont {Icking}},
  \bibinfo {author} {\bibfnamefont {S.}~\bibnamefont
  {M{\ifmmode\ddot{o}\else\"{o}\fi}ller}}, \bibinfo {author} {\bibfnamefont
  {K.}~\bibnamefont {Watanabe}}, \bibinfo {author} {\bibfnamefont
  {T.}~\bibnamefont {Taniguchi}}, \bibinfo {author} {\bibfnamefont
  {C.}~\bibnamefont {Stampfer}}, \ and\ \bibinfo {author} {\bibfnamefont
  {C.}~\bibnamefont {Volk}},\ }\href {https://arxiv.org/abs/2010.14399v1}
  {\bibfield  {journal} {\bibinfo  {journal} {arXiv}\ } (\bibinfo {year}
  {2020}{\natexlab{c}})},\ \Eprint {http://arxiv.org/abs/2010.14399}
  {2010.14399} \BibitemShut {NoStop}%
\bibitem [{\citenamefont {Loss}\ and\ \citenamefont
  {DiVincenzo}(1998)}]{Loss1998Jan}%
  \BibitemOpen
  \bibfield  {author} {\bibinfo {author} {\bibfnamefont {D.}~\bibnamefont
  {Loss}}\ and\ \bibinfo {author} {\bibfnamefont {D.~P.}\ \bibnamefont
  {DiVincenzo}},\ }\href {\doibase 10.1103/PhysRevA.57.120} {\bibfield
  {journal} {\bibinfo  {journal} {Phys. Rev. A}\ }\textbf {\bibinfo {volume}
  {57}},\ \bibinfo {pages} {120} (\bibinfo {year} {1998})}\BibitemShut
  {NoStop}%
\bibitem [{\citenamefont {Watson}\ \emph {et~al.}(2018)\citenamefont {Watson},
  \citenamefont {Philips}, \citenamefont {Kawakami}, \citenamefont {Ward},
  \citenamefont {Scarlino}, \citenamefont {Veldhorst}, \citenamefont {Savage},
  \citenamefont {Lagally}, \citenamefont {Friesen}, \citenamefont
  {Coppersmith}, \citenamefont {Eriksson},\ and\ \citenamefont
  {Vandersypen}}]{Watson2018Feb}%
  \BibitemOpen
  \bibfield  {author} {\bibinfo {author} {\bibfnamefont {T.~F.}\ \bibnamefont
  {Watson}}, \bibinfo {author} {\bibfnamefont {S.~G.~J.}\ \bibnamefont
  {Philips}}, \bibinfo {author} {\bibfnamefont {E.}~\bibnamefont {Kawakami}},
  \bibinfo {author} {\bibfnamefont {D.~R.}\ \bibnamefont {Ward}}, \bibinfo
  {author} {\bibfnamefont {P.}~\bibnamefont {Scarlino}}, \bibinfo {author}
  {\bibfnamefont {M.}~\bibnamefont {Veldhorst}}, \bibinfo {author}
  {\bibfnamefont {D.~E.}\ \bibnamefont {Savage}}, \bibinfo {author}
  {\bibfnamefont {M.~G.}\ \bibnamefont {Lagally}}, \bibinfo {author}
  {\bibfnamefont {M.}~\bibnamefont {Friesen}}, \bibinfo {author} {\bibfnamefont
  {S.~N.}\ \bibnamefont {Coppersmith}}, \bibinfo {author} {\bibfnamefont
  {M.~A.}\ \bibnamefont {Eriksson}}, \ and\ \bibinfo {author} {\bibfnamefont
  {L.~M.~K.}\ \bibnamefont {Vandersypen}},\ }\href {\doibase
  10.1038/nature25766} {\bibfield  {journal} {\bibinfo  {journal} {nature}\
  }\textbf {\bibinfo {volume} {555}},\ \bibinfo {pages} {633} (\bibinfo {year}
  {2018})}\BibitemShut {NoStop}%
\bibitem [{\citenamefont {Yoneda}\ \emph {et~al.}(2017)\citenamefont {Yoneda},
  \citenamefont {Takeda}, \citenamefont {Otsuka}, \citenamefont {Nakajima},
  \citenamefont {Delbecq}, \citenamefont {Allison}, \citenamefont {Honda},
  \citenamefont {Kodera}, \citenamefont {Oda}, \citenamefont {Hoshi},
  \citenamefont {Usami}, \citenamefont {Itoh},\ and\ \citenamefont
  {Tarucha}}]{Yoneda2017Dec}%
  \BibitemOpen
  \bibfield  {author} {\bibinfo {author} {\bibfnamefont {J.}~\bibnamefont
  {Yoneda}}, \bibinfo {author} {\bibfnamefont {K.}~\bibnamefont {Takeda}},
  \bibinfo {author} {\bibfnamefont {T.}~\bibnamefont {Otsuka}}, \bibinfo
  {author} {\bibfnamefont {T.}~\bibnamefont {Nakajima}}, \bibinfo {author}
  {\bibfnamefont {M.~R.}\ \bibnamefont {Delbecq}}, \bibinfo {author}
  {\bibfnamefont {G.}~\bibnamefont {Allison}}, \bibinfo {author} {\bibfnamefont
  {T.}~\bibnamefont {Honda}}, \bibinfo {author} {\bibfnamefont
  {T.}~\bibnamefont {Kodera}}, \bibinfo {author} {\bibfnamefont
  {S.}~\bibnamefont {Oda}}, \bibinfo {author} {\bibfnamefont {Y.}~\bibnamefont
  {Hoshi}}, \bibinfo {author} {\bibfnamefont {N.}~\bibnamefont {Usami}},
  \bibinfo {author} {\bibfnamefont {K.~M.}\ \bibnamefont {Itoh}}, \ and\
  \bibinfo {author} {\bibfnamefont {S.}~\bibnamefont {Tarucha}},\ }\href
  {\doibase 10.1038/s41565-017-0014-x} {\bibfield  {journal} {\bibinfo
  {journal} {Nat. Nanotechnol.}\ }\textbf {\bibinfo {volume} {13}},\ \bibinfo
  {pages} {102} (\bibinfo {year} {2017})}\BibitemShut {NoStop}%
\bibitem [{\citenamefont {Nowack}\ \emph {et~al.}(2011)\citenamefont {Nowack},
  \citenamefont {Shafiei}, \citenamefont {Laforest}, \citenamefont
  {Prawiroatmodjo}, \citenamefont {Schreiber}, \citenamefont {Reichl},
  \citenamefont {Wegscheider},\ and\ \citenamefont
  {Vandersypen}}]{Nowack2011Sep}%
  \BibitemOpen
  \bibfield  {author} {\bibinfo {author} {\bibfnamefont {K.~C.}\ \bibnamefont
  {Nowack}}, \bibinfo {author} {\bibfnamefont {M.}~\bibnamefont {Shafiei}},
  \bibinfo {author} {\bibfnamefont {M.}~\bibnamefont {Laforest}}, \bibinfo
  {author} {\bibfnamefont {G.~E. D.~K.}\ \bibnamefont {Prawiroatmodjo}},
  \bibinfo {author} {\bibfnamefont {L.~R.}\ \bibnamefont {Schreiber}}, \bibinfo
  {author} {\bibfnamefont {C.}~\bibnamefont {Reichl}}, \bibinfo {author}
  {\bibfnamefont {W.}~\bibnamefont {Wegscheider}}, \ and\ \bibinfo {author}
  {\bibfnamefont {L.~M.~K.}\ \bibnamefont {Vandersypen}},\ }\href {\doibase
  10.1126/science.1209524} {\bibfield  {journal} {\bibinfo  {journal}
  {Science}\ }\textbf {\bibinfo {volume} {333}},\ \bibinfo {pages} {1269}
  (\bibinfo {year} {2011})}\BibitemShut {NoStop}%
\bibitem [{\citenamefont {Fujisawa}\ \emph
  {et~al.}(2001{\natexlab{a}})\citenamefont {Fujisawa}, \citenamefont
  {Tokura},\ and\ \citenamefont {Hirayama}}]{Fujisawa2001Feb}%
  \BibitemOpen
  \bibfield  {author} {\bibinfo {author} {\bibfnamefont {T.}~\bibnamefont
  {Fujisawa}}, \bibinfo {author} {\bibfnamefont {Y.}~\bibnamefont {Tokura}}, \
  and\ \bibinfo {author} {\bibfnamefont {Y.}~\bibnamefont {Hirayama}},\ }\href
  {\doibase 10.1103/PhysRevB.63.081304} {\bibfield  {journal} {\bibinfo
  {journal} {Phys. Rev. B}\ }\textbf {\bibinfo {volume} {63}},\ \bibinfo
  {pages} {081304(R)} (\bibinfo {year} {2001}{\natexlab{a}})}\BibitemShut
  {NoStop}%
\bibitem [{\citenamefont {Fujisawa}\ \emph {et~al.}(2002)\citenamefont
  {Fujisawa}, \citenamefont {Austing}, \citenamefont {Tokura}, \citenamefont
  {Hirayama},\ and\ \citenamefont {Tarucha}}]{Fujisawa2002Sep}%
  \BibitemOpen
  \bibfield  {author} {\bibinfo {author} {\bibfnamefont {T.}~\bibnamefont
  {Fujisawa}}, \bibinfo {author} {\bibfnamefont {D.~G.}\ \bibnamefont
  {Austing}}, \bibinfo {author} {\bibfnamefont {Y.}~\bibnamefont {Tokura}},
  \bibinfo {author} {\bibfnamefont {Y.}~\bibnamefont {Hirayama}}, \ and\
  \bibinfo {author} {\bibfnamefont {S.}~\bibnamefont {Tarucha}},\ }\href
  {\doibase 10.1038/nature00976} {\bibfield  {journal} {\bibinfo  {journal}
  {Nature}\ }\textbf {\bibinfo {volume} {419}},\ \bibinfo {pages} {278}
  (\bibinfo {year} {2002})}\BibitemShut {NoStop}%
\bibitem [{\citenamefont {Engels}\ \emph {et~al.}(2014)\citenamefont {Engels},
  \citenamefont {Terr{\ifmmode \acute{e} \else \'{e}\fi}s}, \citenamefont
  {Epping}, \citenamefont {Khodkov}, \citenamefont {Watanabe}, \citenamefont
  {Taniguchi}, \citenamefont {Beschoten},\ and\ \citenamefont
  {Stampfer}}]{Engels2014Sep}%
  \BibitemOpen
  \bibfield  {author} {\bibinfo {author} {\bibfnamefont {S.}~\bibnamefont
  {Engels}}, \bibinfo {author} {\bibfnamefont {B.}~\bibnamefont {Terr{\ifmmode
  \acute{e} \else \'{e}\fi}s}}, \bibinfo {author} {\bibfnamefont
  {A.}~\bibnamefont {Epping}}, \bibinfo {author} {\bibfnamefont
  {T.}~\bibnamefont {Khodkov}}, \bibinfo {author} {\bibfnamefont
  {K.}~\bibnamefont {Watanabe}}, \bibinfo {author} {\bibfnamefont
  {T.}~\bibnamefont {Taniguchi}}, \bibinfo {author} {\bibfnamefont
  {B.}~\bibnamefont {Beschoten}}, \ and\ \bibinfo {author} {\bibfnamefont
  {C.}~\bibnamefont {Stampfer}},\ }\href {\doibase
  10.1103/PhysRevLett.113.126801} {\bibfield  {journal} {\bibinfo  {journal}
  {Phys. Rev. Lett.}\ }\textbf {\bibinfo {volume} {113}},\ \bibinfo {pages}
  {126801} (\bibinfo {year} {2014})}\BibitemShut {NoStop}%
\bibitem [{\citenamefont {McCann}\ and\ \citenamefont
  {Fal{'}ko}(2006)}]{McCann2006Mar}%
  \BibitemOpen
  \bibfield  {author} {\bibinfo {author} {\bibfnamefont {E.}~\bibnamefont
  {McCann}}\ and\ \bibinfo {author} {\bibfnamefont {V.~I.}\ \bibnamefont
  {Fal{'}ko}},\ }\href {\doibase 10.1103/PhysRevLett.96.086805} {\bibfield
  {journal} {\bibinfo  {journal} {Phys. Rev. Lett.}\ }\textbf {\bibinfo
  {volume} {96}},\ \bibinfo {pages} {086805} (\bibinfo {year}
  {2006})}\BibitemShut {NoStop}%
\bibitem [{\citenamefont {Oostinga}\ \emph {et~al.}(2007)\citenamefont
  {Oostinga}, \citenamefont {Heersche}, \citenamefont {Liu}, \citenamefont
  {Morpurgo},\ and\ \citenamefont {Vandersypen}}]{Oos2007Dec}%
  \BibitemOpen
  \bibfield  {author} {\bibinfo {author} {\bibfnamefont {J.~B.}\ \bibnamefont
  {Oostinga}}, \bibinfo {author} {\bibfnamefont {H.~B.}\ \bibnamefont
  {Heersche}}, \bibinfo {author} {\bibfnamefont {X.}~\bibnamefont {Liu}},
  \bibinfo {author} {\bibfnamefont {A.~F.}\ \bibnamefont {Morpurgo}}, \ and\
  \bibinfo {author} {\bibfnamefont {L.~M.~K.}\ \bibnamefont {Vandersypen}},\
  }\href {\doibase 10.1038/nmat2082} {\bibfield  {journal} {\bibinfo  {journal}
  {Nat. Mater.}\ }\textbf {\bibinfo {volume} {7}},\ \bibinfo {pages} {151}
  (\bibinfo {year} {2007})}\BibitemShut {NoStop}%
\bibitem [{\citenamefont {Zhang}\ \emph {et~al.}(2009)\citenamefont {Zhang},
  \citenamefont {Tang}, \citenamefont {Girit}, \citenamefont {Hao},
  \citenamefont {Martin}, \citenamefont {Zettl}, \citenamefont {Crommie},
  \citenamefont {Shen},\ and\ \citenamefont {Wang}}]{Zhang2009Jun}%
  \BibitemOpen
  \bibfield  {author} {\bibinfo {author} {\bibfnamefont {Y.}~\bibnamefont
  {Zhang}}, \bibinfo {author} {\bibfnamefont {T.-T.}\ \bibnamefont {Tang}},
  \bibinfo {author} {\bibfnamefont {C.}~\bibnamefont {Girit}}, \bibinfo
  {author} {\bibfnamefont {Z.}~\bibnamefont {Hao}}, \bibinfo {author}
  {\bibfnamefont {M.~C.}\ \bibnamefont {Martin}}, \bibinfo {author}
  {\bibfnamefont {A.}~\bibnamefont {Zettl}}, \bibinfo {author} {\bibfnamefont
  {M.~F.}\ \bibnamefont {Crommie}}, \bibinfo {author} {\bibfnamefont {Y.~R.}\
  \bibnamefont {Shen}}, \ and\ \bibinfo {author} {\bibfnamefont
  {F.}~\bibnamefont {Wang}},\ }\href {\doibase 10.1038/nature08105} {\bibfield
  {journal} {\bibinfo  {journal} {Nature}\ }\textbf {\bibinfo {volume} {459}},\
  \bibinfo {pages} {820} (\bibinfo {year} {2009})}\BibitemShut {NoStop}%
\bibitem [{\citenamefont {Garreis}\ \emph {et~al.}(2020)\citenamefont
  {Garreis}, \citenamefont {Knothe}, \citenamefont {Tong}, \citenamefont
  {Eich}, \citenamefont {Gold}, \citenamefont {Watanabe}, \citenamefont
  {Taniguchi}, \citenamefont {Fal'ko}, \citenamefont {Ihn}, \citenamefont
  {Ensslin},\ and\ \citenamefont {Kurzmann}}]{Garreis2020Nov}%
  \BibitemOpen
  \bibfield  {author} {\bibinfo {author} {\bibfnamefont {R.}~\bibnamefont
  {Garreis}}, \bibinfo {author} {\bibfnamefont {A.}~\bibnamefont {Knothe}},
  \bibinfo {author} {\bibfnamefont {C.}~\bibnamefont {Tong}}, \bibinfo {author}
  {\bibfnamefont {M.}~\bibnamefont {Eich}}, \bibinfo {author} {\bibfnamefont
  {C.}~\bibnamefont {Gold}}, \bibinfo {author} {\bibfnamefont {K.}~\bibnamefont
  {Watanabe}}, \bibinfo {author} {\bibfnamefont {T.}~\bibnamefont {Taniguchi}},
  \bibinfo {author} {\bibfnamefont {V.}~\bibnamefont {Fal'ko}}, \bibinfo
  {author} {\bibfnamefont {T.}~\bibnamefont {Ihn}}, \bibinfo {author}
  {\bibfnamefont {K.}~\bibnamefont {Ensslin}}, \ and\ \bibinfo {author}
  {\bibfnamefont {A.}~\bibnamefont {Kurzmann}},\ }\href
  {https://arxiv.org/abs/2011.07951v1} {\bibfield  {journal} {\bibinfo
  {journal} {arXiv}\ } (\bibinfo {year} {2020})},\ \Eprint
  {http://arxiv.org/abs/2011.07951} {2011.07951} \BibitemShut {NoStop}%
\bibitem [{\citenamefont {Banszerus}\ \emph
  {et~al.}(2020{\natexlab{d}})\citenamefont {Banszerus}, \citenamefont
  {Fabian}, \citenamefont {M{\ifmmode\ddot{o}\else\"{o}\fi}ller}, \citenamefont
  {Icking}, \citenamefont {Heiming}, \citenamefont {Trellenkamp}, \citenamefont
  {Lentz}, \citenamefont {Neumaier}, \citenamefont {Otto}, \citenamefont
  {Watanabe}, \citenamefont {Taniguchi}, \citenamefont {Libisch}, \citenamefont
  {Volk},\ and\ \citenamefont {Stampfer}}]{Banszerus2020Sep}%
  \BibitemOpen
  \bibfield  {author} {\bibinfo {author} {\bibfnamefont {L.}~\bibnamefont
  {Banszerus}}, \bibinfo {author} {\bibfnamefont {T.}~\bibnamefont {Fabian}},
  \bibinfo {author} {\bibfnamefont {S.}~\bibnamefont
  {M{\ifmmode\ddot{o}\else\"{o}\fi}ller}}, \bibinfo {author} {\bibfnamefont
  {E.}~\bibnamefont {Icking}}, \bibinfo {author} {\bibfnamefont
  {H.}~\bibnamefont {Heiming}}, \bibinfo {author} {\bibfnamefont
  {S.}~\bibnamefont {Trellenkamp}}, \bibinfo {author} {\bibfnamefont
  {F.}~\bibnamefont {Lentz}}, \bibinfo {author} {\bibfnamefont
  {D.}~\bibnamefont {Neumaier}}, \bibinfo {author} {\bibfnamefont
  {M.}~\bibnamefont {Otto}}, \bibinfo {author} {\bibfnamefont {K.}~\bibnamefont
  {Watanabe}}, \bibinfo {author} {\bibfnamefont {T.}~\bibnamefont {Taniguchi}},
  \bibinfo {author} {\bibfnamefont {F.}~\bibnamefont {Libisch}}, \bibinfo
  {author} {\bibfnamefont {C.}~\bibnamefont {Volk}}, \ and\ \bibinfo {author}
  {\bibfnamefont {C.}~\bibnamefont {Stampfer}},\ }\href {\doibase
  10.1002/pssb.202000333} {\bibfield  {journal} {\bibinfo  {journal} {Phys.
  Status Solidi B}\ }\textbf {\bibinfo {volume} {n/a}},\ \bibinfo {pages}
  {2000333} (\bibinfo {year} {2020}{\natexlab{d}})}\BibitemShut {NoStop}%
\bibitem [{\citenamefont {Volk}\ \emph {et~al.}(2013)\citenamefont {Volk},
  \citenamefont {Neumann}, \citenamefont {Kazarski}, \citenamefont {Fringes},
  \citenamefont {Engels}, \citenamefont {Haupt}, \citenamefont {M{\ifmmode
  \ddot{u} \else \"{u}\fi}ller},\ and\ \citenamefont {Stampfer}}]{Volk2013Apr}%
  \BibitemOpen
  \bibfield  {author} {\bibinfo {author} {\bibfnamefont {C.}~\bibnamefont
  {Volk}}, \bibinfo {author} {\bibfnamefont {C.}~\bibnamefont {Neumann}},
  \bibinfo {author} {\bibfnamefont {S.}~\bibnamefont {Kazarski}}, \bibinfo
  {author} {\bibfnamefont {S.}~\bibnamefont {Fringes}}, \bibinfo {author}
  {\bibfnamefont {S.}~\bibnamefont {Engels}}, \bibinfo {author} {\bibfnamefont
  {F.}~\bibnamefont {Haupt}}, \bibinfo {author} {\bibfnamefont
  {A.}~\bibnamefont {M{\ifmmode \ddot{u} \else \"{u}\fi}ller}}, \ and\ \bibinfo
  {author} {\bibfnamefont {C.}~\bibnamefont {Stampfer}},\ }\href {\doibase
  10.1038/ncomms2738} {\bibfield  {journal} {\bibinfo  {journal} {Nat.
  Commun.}\ }\textbf {\bibinfo {volume} {4}},\ \bibinfo {pages} {1753}
  (\bibinfo {year} {2013})}\BibitemShut {NoStop}%
\bibitem [{\citenamefont {Fujisawa}\ \emph
  {et~al.}(2001{\natexlab{b}})\citenamefont {Fujisawa}, \citenamefont
  {Tokura},\ and\ \citenamefont {Hirayama}}]{Fujisawa2001Apr}%
  \BibitemOpen
  \bibfield  {author} {\bibinfo {author} {\bibfnamefont {T.}~\bibnamefont
  {Fujisawa}}, \bibinfo {author} {\bibfnamefont {Y.}~\bibnamefont {Tokura}}, \
  and\ \bibinfo {author} {\bibfnamefont {Y.}~\bibnamefont {Hirayama}},\ }\href
  {\doibase 10.1016/S0921-4526(01)00385-4} {\bibfield  {journal} {\bibinfo
  {journal} {Physica B}\ }\textbf {\bibinfo {volume} {298}},\ \bibinfo {pages}
  {573} (\bibinfo {year} {2001}{\natexlab{b}})}\BibitemShut {NoStop}%
\bibitem [{\citenamefont {Hanson}\ \emph {et~al.}(2003)\citenamefont {Hanson},
  \citenamefont {Witkamp}, \citenamefont {Vandersypen}, \citenamefont {van
  Beveren}, \citenamefont {Elzerman},\ and\ \citenamefont
  {Kouwenhoven}}]{Hanson2003Nov}%
  \BibitemOpen
  \bibfield  {author} {\bibinfo {author} {\bibfnamefont {R.}~\bibnamefont
  {Hanson}}, \bibinfo {author} {\bibfnamefont {B.}~\bibnamefont {Witkamp}},
  \bibinfo {author} {\bibfnamefont {L.~M.~K.}\ \bibnamefont {Vandersypen}},
  \bibinfo {author} {\bibfnamefont {L.~H.~W.}\ \bibnamefont {van Beveren}},
  \bibinfo {author} {\bibfnamefont {J.~M.}\ \bibnamefont {Elzerman}}, \ and\
  \bibinfo {author} {\bibfnamefont {L.~P.}\ \bibnamefont {Kouwenhoven}},\
  }\href {\doibase 10.1103/PhysRevLett.91.196802} {\bibfield  {journal}
  {\bibinfo  {journal} {Phys. Rev. Lett.}\ }\textbf {\bibinfo {volume} {91}},\
  \bibinfo {pages} {196802} (\bibinfo {year} {2003})}\BibitemShut {NoStop}%
\bibitem [{\citenamefont {Churchill}\ \emph {et~al.}(2009)\citenamefont
  {Churchill}, \citenamefont {Bestwick}, \citenamefont {Harlow}, \citenamefont
  {Kuemmeth}, \citenamefont {Marcos}, \citenamefont {Stwertka}, \citenamefont
  {Watson},\ and\ \citenamefont {Marcus}}]{Churchill2009Apr}%
  \BibitemOpen
  \bibfield  {author} {\bibinfo {author} {\bibfnamefont {H.~O.~H.}\
  \bibnamefont {Churchill}}, \bibinfo {author} {\bibfnamefont {A.~J.}\
  \bibnamefont {Bestwick}}, \bibinfo {author} {\bibfnamefont {J.~W.}\
  \bibnamefont {Harlow}}, \bibinfo {author} {\bibfnamefont {F.}~\bibnamefont
  {Kuemmeth}}, \bibinfo {author} {\bibfnamefont {D.}~\bibnamefont {Marcos}},
  \bibinfo {author} {\bibfnamefont {C.~H.}\ \bibnamefont {Stwertka}}, \bibinfo
  {author} {\bibfnamefont {S.~K.}\ \bibnamefont {Watson}}, \ and\ \bibinfo
  {author} {\bibfnamefont {C.~M.}\ \bibnamefont {Marcus}},\ }\href {\doibase
  10.1038/nphys1247} {\bibfield  {journal} {\bibinfo  {journal} {Nat. Phys.}\
  }\textbf {\bibinfo {volume} {5}},\ \bibinfo {pages} {321} (\bibinfo {year}
  {2009})}\BibitemShut {NoStop}%
\bibitem [{\citenamefont {Albrecht}\ \emph {et~al.}(2017)\citenamefont
  {Albrecht}, \citenamefont {Moers},\ and\ \citenamefont
  {Hermanns}}]{Albrecht2017May}%
  \BibitemOpen
  \bibfield  {author} {\bibinfo {author} {\bibfnamefont {W.}~\bibnamefont
  {Albrecht}}, \bibinfo {author} {\bibfnamefont {J.}~\bibnamefont {Moers}}, \
  and\ \bibinfo {author} {\bibfnamefont {B.}~\bibnamefont {Hermanns}},\ }\href
  {\doibase 10.17815/jlsrf-3-158} {\bibfield  {journal} {\bibinfo  {journal}
  {Journal of Large-Scale Research Facilities}\ }\textbf {\bibinfo {volume}
  {3}},\ \bibinfo {pages} {112} (\bibinfo {year} {2017})}\BibitemShut {NoStop}%
\end{thebibliography}%

\end{document}